\documentclass[useAMS,usenatbib]{mn2e}
\usepackage{epsfig}
\usepackage{color}
\usepackage{subfigure}

\def\lsim{\mathrel{\rlap{\lower3.5pt\hbox{\hskip0.5pt$\sim$}}
    \raise0.5pt\hbox{$<$}}}
\def\gsim{~\rlap{$>$}{\lower 1.0ex\hbox{$\sim$}}}

\newcommand{\goodgap}{\hspace{\subfigtopskip} \hspace{\subfigbottomskip}}

\title[Dark matter haloes properties in dwarfs galaxies]{Statistical properties of the dark matter haloes of dwarfs galaxies and correlations with the environment}

\author[A. Del Popolo \& V.F. Cardone]{A. Del Popolo$^{1}$, V.F. Cardone$^{2}$\footnote{Corresponding author\,: {\tt winnyenodrac@gmail.com}} \\
$^1$Dipartimento di Fisica e Astronomia, Universitˆ di Catania, Viale Andrea Doria 6, 95125 Catania, Italy \\
$^2$I.N.A.F. - Osservatorio Astronomico di Roma, via Frascati 33, 00040\,-\,Monte Porzio Catone (Roma), Italy \\}

\date{Accepted xxx, Received yyy, in original form zzz}

\begin{document}

\maketitle

\begin{abstract}

According to the now strongly supported concordance $\Lambda$CDM model, galaxies may be grossly described as a luminous component embedded in a dark matter halo. The density profile of these mass dominating haloes may be determined by N\,-\,body simulations which mimic the evolution of the tiny initial density perturbations during the process leading to the structures we observe today. Unfortunately, when the effect of baryons is taken into account, the situation gets much more complicated due to the difficulties in simulating their physics. As a consequence, a definitive prediction of how dark matter haloes should presently look like is still missing. We revisit here this issue from an observational point of view devoting our attention to dwarf galaxies. Being likely dark matter dominated, these systems are ideal candidates to investigate the present day halo density profiles and check whether dark matter related quantities correlate with the stellar ones or the environment. By fitting a large sample of well measured rotation curves, we infer constraints on both halo structural parameters (such as the logarithmic slope of the density profile and its concentration) and derived quantities (e.g., the mass fraction and the Newtonian acceleration) which could then be used to constrain galaxy formation scenarios. Moreover, we investigate whether the halo properties correlates with the environment the galaxy lives in thus offering a new tool to deepen our understanding of galaxy formation.

\end{abstract}


\section{Introduction}

The discovery of the cosmic speed up through the Hubble diagram of Type Ia Supernovae more than a decade ago has pointed at the cosmological constant $\Lambda$ as the ideal candidate to drive the accelerated expansion. Almost fifteen years later, the $\Lambda$CDM model, made out by the $\Lambda$ term dominating the energy budget and the cold dark matter (CDM) being responsible for the clustering on galaxy and clusters scales, has proved to excellently work in fitting a wide variety of data on cosmological scales, running from SNeIa \citep{Union2} to Baryon Acoustic Oscillations \citep{P10} and cosmic microwave background anisotropy and polarization spectra \citep{WMAP7}.

The unprecedented accuracy in describing the large scale universe is frustratingly contrasted by the serious shortcomings the $\Lambda$CDM model has to deal with on the galaxy scales, the most well known difficulties being represented by the substructure problem, i.e. the strong discrepancy between the predicted and observed number of satellites \citep{Kl99,Moo99,sub}, and the cusp/core controversy \citep{FP94,Moo94,deB10} with rotation curves data arguing in favour of cored density profiles instead of the cusped ones coming out of dissipationless $\Lambda$CDM based N\-\,body simulations. Notwithstanding the significant progresses in both modeling and observations during the last 15 years, both these problems are still on the ground, thus motivating further investigations.

To this end, dwarfs galaxies stand out as ideal candidates to make our knowledge of haloes a step forward. These small mass systems, characterized by having low values of luminosity, metallicity and size, are important building blocks of more massive galaxies. In CDM cosmologies, the smallest dwarf galaxies, likely formed before reionization with masses smaller than $10^8$\,-\,$10^9 \ {\rm M_{\odot}}$ (see, e.g., \citealt{GO97}).

Morphologically, dwarf galaxies can be classified as dwarf spiral galaxies, blue compact dwarf galaxies (BCDs), dwarf irregular galaxies (dIrrs), dwarf elliptical galaxies (dEs), dwarf spheroidal galaxies (dSphs), ultra\,-\,faint dwarf spheroidals, and tidal dwarf galaxies. Dwarf spiral galaxies, BCDs and dIrrs are star\,-\,forming objects, gas rich and rotationally supported, while, on the contrary, late\,-\,type dwarf spirals are slow rotators or exhibit solid\,-\,body rotation. dEs usually have little or no detectable gas, and are often not rotationally supported, similarly to dSphs. Tidal dwarf galaxies, finally, form from the debris torn out of more massive galaxies during interactions and mergers.

Similarly interesting hints on the cusp/core controversy may be drawn from Low Surface Brightness (LSB) galaxies. These are late\,-\,type, gas\,-\,rich, dark matter dominated disk galaxies. Their optical appearance is dominated by an exponential disk with a young, blue population, with little evidence for a dominant old population. Additionally, these galaxies do not have large dominant bulges\footnote{For precision's sake, another type of LSB galaxies often discussed in the literature are the massive, early\,-\,type, bulge\,-\,dominated LSB galaxies. These galaxies have properties entirely different from the late\,-\,type LSB galaxies \citep{Spray95,Pick97}.}.

Remarkably, whether the dwarfs haloes are cored or cuspy was still a matter of controversy till recent times. On one hand, many studies concordantly pointed to cored profile as a better fit to the rotation curve data \citep{B95,K98,BS01,M02,S05}. On the contrary, some further works seemed to contradict this conclusion with the fit to high resolution data being equally good for both cored and steeper profiles (se, e.g., \citealt{Sw03}). Although observational problems (such as slit offsets and beam smearing) and theoretical uncertainties (related to non circular motions and triaxiality) were originally thought to bias the determination of the inner slopes of the density profiles from rotation curves data (see, e.g., \citealt{Sp05}), recent studies (e.g., \citealt{dNK11}) have convincingly shown that this is not the case so that one can safely rely on measured circular velocity profiles to investigate the cusp/core controversy.

It is worth noting that cuspy haloes are predicted from dark matter only simulations \citep{NFW96,NFW97,Moore+98,P03,N04,N10} so that the impact of baryons, which actually dominate the inner regions, is incorrectly neglected. It is interesting to note that the first solution proposed to the cusp/core problem is connected to baryon physics. \cite{NEF96} indeed studied how supernova\,-\,driven winds that expel a large fraction of baryons in the halo could give rise to a dark matter core. In the same framework, \cite{GSL99} reproduced the observed rotation curve of DDO 154 by simulating NFW haloes and exposing them to violent gas outflows, while \cite{GZ02} considered the maximal limit of complete blow out of all of the baryons. \cite{RG05} then refined the analysis showing that, in order the previous mechanism to be efficient, it is necessary to have two impulsive mass\,-\,loss phases punctuated by gas re\,-\,accretion. More recent cosmological simulations \citep{Masch06,Gov10} have continued to study the effects of baryons on the inner DM inner profile. In particular, \cite{Gov10} have recently showed that strong outflows from supernovae explosions in dwarfs\,-\,like systems may remove the low angular momentum gas inhibiting the formation of bulges and leading to shallow central dark matter profiles. Such a mechanism should, however, be inefficient in systems with a deeper potential well such as LSBs where the cusp/core problem is nevertheless present with several works arguing in favour of both cored \citep{dBB02} and cuspy \citep{vdB00} profiles. This circumstance has prompted the interest for alternative solutions to rescue the $\Lambda$CDM paradigm without drastic changes to its physics. An incomplete list contains interactions of the DM with a stellar bar \citep{WK02,McMD05}, decay of binary black hole orbits after galaxies merge \citep{MM01}, baryon energy feedback from active galactic nucleus \citep{Pei08}, dynamical friction of stellar/DM clumps against the background halo \citep{ElZ01,ElZ04,RD08,RD09}.

A possible way out of the cusp/core problem is hidden in the recent result of de Blok et al. (2008). Fitting the high resolution rotation curve data of the THINGS galaxy sample \citep{W08}, \cite{Things} have found that a NFW profile or a cored isothermal model fit equally well the data for galaxies with $M_B \leq -19$, while cuspy profiles are strongly rejected for systems with $M_B \geq -19$. In other terms, for low mass galaxies, a core dominated halo is clearly preferred over a cusp\,-\,like one, while both models are equally allowed for massive disk dominated systems. That the density profile of dark matter haloes is not universal has been actually also suggested by recent numerical simulations \citep{JS00,R03,N04,Gr06,Mer06,N10} showing that departures from the canonical NFW profile at $1$\,-\,$10\%$ of the virial radius. Moreover, \cite{DeP09,DeP10} used an analytical approach to show that such departures are stronger when baryons are taken into account with the inner slopes of the density profile becoming mass and redshift dependent (lower mass haloes having shallower profiles, with the slope becoming steeper towards higher $z$).

Abandoning the universality of the halo density profile opens up the hunt for what is responsible for the variety of the halo properties which are inferred from observations. In order to shed light on this issue, it is worth looking at the statistical distribution of halo parameters and dark matter related quantities also looking for correlations with both the stellar properties and the environment the halo is embedded in. Motivated by this consideration, we have here analyzed the rotation curves of a sample of 27 dwarfs and 10 LSB galaxies collected from the literature. All of them have well measured rotation curves with both HI and H$\alpha$ data extending to large $R/R_d$ values (with $R_d$ the disk scalelength) so that one is able to probe the dark matter dominated regions. Moreover, the stellar contribution to the circular velocity is well constrained by the measured luminosity profile and the constraints on the stellar population making it possible to constrain the stellar mass\,-\,to\,-\,light ratio and hence the baryon matter distribution.

The scheme of the paper is as follows. The halo model parametrization and the method used to constraint its parameters are presented in Sect. 2, while the results of the rotation curve fits are discussed in Sect. 3. We then investigate the correlation of different halo properties with environment indicators in Sect. 4 and devote Sect. 5 to conclusions.

\begin{table*}
%\tiny
%\scriptsize
\caption{Galaxy sample and constraints on model parameters. Columns are as follows\,: 1. galaxy id; 2. type (0 for dwarfs, 1 for LSBs); 3. reference for the rotation curve data according to the following scheme\,: Oh = Oh et al. (2011), dBB02 = de Blok \& Bosma (1992), THINGS = de Blok et al. (2008), S05 = Simon et al. (2005),  Sw11 = Swaters et al. (2011); 4. tidal index; 5. projected distance of the tenth closest galaxy (in kpc); 6. number of galaxies within a projected distance of 750 kpc; 7.\,-\,11. median and $68\%$ confidence range for the model parameters $(\Upsilon/\Upsilon_{fid}, \alpha, c_{vir}, V_{vir}, \log{M_{vir}})$ assuming an Einasto density profile for the DM halo; 12. reduced $\tilde{\chi}^2 = \chi^2/d.o.f.$ for the best fit model. Note that, for galaxies in the S05 sample, we have to set the stellar $M/L$ ratio to its fiducial value since we have only data on the DM circular velocity so that no constraints on $\Upsilon_{\star}/\Upsilon_{fid} = 1.0$ can be reported in the table.}
\begin{center}
\begin{tabular}{cccccccccccc}
\hline
Id & Type & Ref & $\Theta$ & $R_{10}$ & ${\cal{N}}_{750}$ & $\Upsilon_{\star}/\Upsilon_{fid}$ & $n_{DM}$ & $c_{vir}$ & $V_{vir}$ & $\log{M_{vir}}$ & $\tilde{\chi}^2$ \\
\hline
DDO53 & 0 &	Oh & 0.7 & 411.724 & 24 & $0.81_{-0.19}^{+0.37}$ & $1.29_{-0.83}^{+1.17}$ & $8.1_{-2.5}^{+1.6}$ & $27.5_{-13.2}^{+45.6}$ & $10.0_{-0.8}^{+1.3}$ & 0.19 \\
DDO185 & 0 & dBB02 & $\ldots$ & 342.667 & 26 & $0.80_{-0.18}^{+1.13}$ & $0.63_{-0.46}^{+1.13}$ & $11.5_{-2.4}^{+1.1}$ & $56.1_{-30.8}^{+100.1}$ & $10.9_{-1.0}^{+1.3}$ & 1.45 \\
HoI & 0 & Oh & 1.5 & 187.429 & 27 & $0.75_{-0.14}^{+0.29}$ & $0.29_{-0.08}^{+0.09}$ & $18.6_{-0.7}^{+0.6}$ & $8.4_{-0.2}^{+0.2}$ &
$8.4_{-0.1}^{+0.1}$ & 2.18 \\
HoII & 0 & Oh & 0.6 & 445.606 & 17 & $0.74_{-0.14}^{+0.29}$ & $12.4_{-4.1}^{+4.3}$ & $3.6_{-2.2}^{+2.8}$ & $29.8_{-3.1}^{+4.1}$ &
$10.1_{-0.2}^{+0.2}$ & 2.37 \\
IC2233 & 1 & dBB02 & $\ldots$ & 20.943 & 8 & $0.91_{-0.28}^{+0.37}$ & $2.00_{-0.94}^{+1.73}$ & $7.1_{-3.5}^{+2.4}$ & $134.2_{-75.0}^{+185.4}$ & $12.1_{-1.1}^{+1.1}$ & 0.70 \\
IC2574 & 0 & THINGS & 0.9 & 172.046 & 26 & $0.76_{-0.14}^{+0.20}$ & $1.17_{-0.47}^{+0.73}$ & $6.3_{-1.5}^{+0.7}$ & $62.9_{-21.1}^{+57.1}$ & $11.1_{-0.6}^{+0.8}$ & 0.50 \\
M81dwB & 0 & Oh & -0.9 & 119.732 & 21 & $0.86_{-0.23}^{+0.37}$ & $2.64_{-1.45}^{+3.88}$ & $24.9_{-14.3}^{+4.9}$ & $17.4_{-7.1}^{+32.5}$ & $9.4_{-0.7}^{+1.4}$ & 0.08 \\	
NGC2366 & 0 & THINGS & 1.0 & 525.228 & 10 & $0.84_{-0.22}^{+0.39}$ & $1.04_{-0.30}^{+0.47}$ & $12.8_{-1.1}^{+0.7}$ & $21.9_{-2.7}^{+5.5}$ & $9.7_{-0.2}^{+0.3}$ & 0.50 \\
NGC2976	& 0	& THINGS & 2.7 & 131.728 & 27 & $0.70_{-0.11}^{+0.25}$ & $0.30_{-0.21}^{+0.35}$ & $10.8_{-0.9}^{+0.6}$ & $69.0_{-33.0}^{+124.4}$ & $11.2_{-0.9}^{+1.3}$ & 0.50 \\
NGC3274	& 1 & dBB02 & -0.3 & 310.581 & 14 & $0.77_{-0.17}^{+0.38}$ & $2.08_{-0.42}^{+0.57}$ & $28.6_{-2.2}^{+2.1}$ & $29.4_{-1.9}^{+1.7}$ & $10.1_{-0.2}^{+0.1}$ & 0.49 \\
NGC4605	& 0 & S05 &	-1.1 & 417.435 & 18 & $\ldots$ & $0.60_{-0.09}^{+0.17}$ & $28.5_{-0.6}^{+0.4}$ & $19.9_{-1.1}^{+1.7}$ &
$9.6_{-0.1}^{+0.1}$ & 1.37 \\
NGC5023	& 1 & dBB02 & -0.5 & 210.311 & 24 & $0.76_{-0.15}^{+0.66}$ & $1.00_{-0.30}^{+0.42}$ & $18.3_{-1.2}^{+1.0}$  & $29.5_{-2.5}^{+4.7}$ & $10.1_{-0.1}^{+0.2}$ & 0.19 \\
NGC5949	& 0	& S05 & $\ldots$ & 479.538 & 3 & $\ldots$ & $3.98_{-2.48}^{+2.85}$ & $8.9_{-6.3}^{+9.0}$ & $129.9_{-93.1}^{+190.3}$ &
$12.1_{-1.7}^{+1.1}$ & 0.25 \\
NGC5963 & 0	& S05 & $\ldots$ & 314.317 & 6 & $\ldots$ & $9.1_{-3.3}^{+3.2}$ & $4.7_{-3.3}^{+8.9}$ & $222.6_{-94.9}^{+102.8}$ &
$12.7_{-0.7}^{+0.5}$ & 0.42 \\
NGC6689	& 0	& S05 & $\ldots$ & $\ldots$ & 2 & $\ldots$ & $3.27_{-1.27}^{+2.22}$ & $9.4_{-5.9}^{+4.1}$ & $150.6_{-74.2}^{+231.1}$ & $12.2_{-0.9}^{+1.2}$ & 0.31 \\
UGC731 & 0 & Sw11 & $\ldots$ & $\ldots$ & 1	& $0.83_{-0.21}^{+0.37}$ & $3.79_{-1.33}^{+2.99}$ & $15.8_{-7.0}^{+2.6}$ & $46.2_{-12.0}^{+28.2}$ & $10.7_{-0.4}^{+0.6}$ & 0.11 \\
UGC1230	& 1 & dBB02 & $\ldots$ & 142.162 & 2 & $0.83_{-0.21}^{+0.37}$ & $1.80_{-0.56}^{+0.81}$ & $14.3_{-1.3}^{+1.7}$ & $52.7_{-4.1}^{+6.8}$ & $10.8_{-0.1}^{+0.2}$ & 0.50 \\
UGC1281	& 1 & dBB02 & -1.2 & 324.170 & 17 & $0.71_{-0.12}^{+0.21}$ & $0.54_{-0.30}^{+0.53}$ & $11.4_{-1.1}^{+0.6}$ & $22.6_{-3.5}^{+11.2}$ & $9.7_{-0.2}^{+0.6}$ & 0.07 \\
UGC3137	& 1	& dBB02 & $\ldots$ & 429.129 & 0 & $0.61_{-0.04}^{+0.11}$ & $0.52_{-0.07}^{+0.10}$ & $8.2_{-0.3}^{+0.2}$ & $49.3_{-0.9}^{+0.7}$ & $10.7_{-0.1}^{+0.1}$ & 2.92 \\
UGC3371	& 1	& dBB02 & $\ldots$ & 517.837 & 1 & $0.79_{-0.17}^{+0.35}$ & $1.49_{-0.65}^{+0.86}$ & $9.1_{-1.7}^{+1.4}$ & $52.9_{-14.9}^{+29.3}$ & $10.8_{-0.4}^{+0.6}$ & 0.01 \\
UGC4173	& 0	& dBB02 & $\ldots$ & 587.408 & 1 & $0.84_{-0.22}^{+0.37}$ & $2.01_{-1.03}^{+1.42}$ & $4.8_{-2.0}^{+1.5}$ & $42.2_{-14.0}^{+34.3}$ & $10.7_{-0.7}^{+0.6}$ & 0.06 \\
UGC4325	& 1	& dBB02 & $\ldots$ & 315.034 & 4 & $0.84_{-0.20}^{+0.39}$ & $1.19_{-0.60}^{+0.77}$ & $16.0_{-3.0}^{+3.9}$ & $86.8_{-36.7}^{+107.4}$ & $11.5_{-0.7}^{+1.0}$ & 0.01 \\
UGC4499	& 0 & Sw11 & $\ldots$ & 374.473 & 6 & $0.83_{-0.21}^{+0.35}$ & $2.58_{-1.04}^{+2.90}$ & $11.3_{-5.2}^{+1.8}$ & $44.2_{-11.5}^{+35.1}$ &
$10.6_{-0.4}^{+0.8}$ & 0.33 \\
UGC5414 & 0 & Sw11 & $\ldots$ & 402.487 & 1	& $0.86_{-0.23}^{+0.37}$ & $2.38_{-0.99}^{+1.93}$ & $9.4_{-4.5}^{+2.7}$ & $50.3_{-21.1}^{+55.2}$ &
$10.8_{-0.7}^{+0.9}$ & 0.12 \\
UGC6446	& 0 & Sw11 & $\ldots$ & 173.865 & 15 & $0.96_{-0.30}^{+0.34}$ & $3.20_{-0.93}^{+2.40}$ & $15.6_{-4.2}^{+1.8}$ & $42.9_{-7.7}^{+17.7}$ &
$10.6_{-0.3}^{+0.4}$ & 0.22 \\
UGC7323 & 0	& Sw11 & $\ldots$ & 93.111 & 67 & $0.91_{-0.27}^{+0.32}$ & $2.30_{-0.90}^{+1.87}$ & $9.2_{-4.7}^{+2.5}$ & $76.7_{-31.8}^{+107.6}$ &
$11.3_{-0.7}^{+1.2}$ & 0.22 \\
UGC7399	& 0	& Sw11 & $\ldots$ & 87.709 & 53	& $1.03_{-0.35}^{+0.29}$ & $4.84_{-0.97}^{+1.36}$ & $20.8_{-2.6}^{+1.8}$ & $65.4_{-8.1}^{+10.8}$ &
$11.1_{-0.2}^{+0.2}$ & 1.55 \\
UGC7524	& 0	& Sw11 & 0.1 & 160.137 & 142 & $0.92_{-0.28}^{+0.36}$ & $2.04_{-0.50}^{+1.23}$ & $14.0_{-2.4}^{+0.9}$ & $40.0_{-6.0}^{+17.0}$ &
$10.5_{-0.2}^{+0.4}$ & 0.18 \\
UGC7559	& 0	& Sw11 & 0.1 & 169.059 & 83 & $0.84_{-0.22}^{+0.35}$ & $2.25_{-1.13}^{+3.01}$ & $14.6_{-7.7}^{+3.4}$ & $19.1_{-7.4}^{+27.4}$ &
$9.5_{-0.6}^{+1.2}$ & 0.03 \\
UGC7603	& 0	& Sw11 & $\ldots$ & 228.234 & 144 & $0.95_{-0.30}^{+0.33}$ & $1.82_{-0.66}^{+1.13}$ & $14.6_{-2.5}^{+1.3}$ & $29.9_{-5.7}^{+13.3}$ &
$10.1_{-0.3}^{+0.5}$ & 0.17 \\
UGC8490 & 0 & Sw11 & -1.1 & 413.572 & 10 & $0.83_{-0.21}^{+0.39}$ & $4.08_{-0.82}^{+1.21}$ & $25.5_{-2.5}^{+2.2}$ & $40.7_{-3.9}^{+5.7}$ &
$10.5_{-0.1}^{+0.2}$ & 0.11 \\
UGC9211	& 0	& Sw11 & $\ldots$ & 395.074 & 2 & $0.83_{-0.21}^{+0.34}$ & $2.22_{-0.99}^{+3.45}$ & $14.4_{-5.6}^{+1.7}$ & $32.9_{-6.9}^{+28.0}$ & $10.2_{-0.3}^{+0.8}$ & 0.03 \\
UGC10310 & 1 & dBB02 & $\ldots$ & $\ldots$ & 0 & $0.82_{-0.20}^{+0.33}$ & $1.71_{-0.93}^{+1.42}$ & $8.2_{-3.2}^{+2.1}$ & $49.7_{-19.2}^{+61.1}$ & $10.8_{-0.7}^{+1.0}$ & 0.09 \\
UGC11707 & 0 & Sw11 & $\ldots$ & $\ldots$ & 3 & $0.83_{-0.21}^{+0.33}$ & $6.65_{-3.36}^{+4.74}$ & $8.2_{-4.5}^{+4.0}$ & $82.6_{-24.0}^{+20.4}$ &
$11.4_{-0.4}^{+0.3}$ & 0.47 \\
UGC12060 & 0 & Sw11 & $\ldots$ & 447.109 & 1 & $0.86_{-0.23}^{+0.36}$ & $4.36_{-2.16}^{+8.29}$ & $21.7_{-9.0}^{+4.1}$ & $39.5_{-9.0}^{+24.7}$ &
$10.5_{-0.4}^{+0.6}$ & 0.05 \\
UGC12632 & 0 & Sw11 & $\ldots$ & $\ldots$ & 2 & $0.82_{-0.18}^{+0.38}$ & $3.47_{-1.35}^{+4.40}$ & $15.4_{-7.7}^{+1.7}$ & $44.1_{-10.3}^{+30.1}$ &
$10.6_{-0.3}^{+0.7}$ & 0.21 \\
UGC12732 & 0 & Sw11 & $\ldots$ & $\ldots$ & 0 & $0.90_{-0.26}^{+0.34}$ & $6.10_{-2.11}^{+2.80}$ & $6.6_{-3.3}^{+3.7}$ & $87.1_{-20.8}^{+42.6}$ & $11.5_{-0.4}^{+0.3}$ & 0.21 \\
\hline
\end{tabular}
\end{center}
\label{tab: fitres}
\end{table*}

\section{Data and method}

Being, by definition, not observable, the dark matter haloes can be probed only indirectly by their effect on the galaxy dynamics or through their lensing effect. Fitting the rotation curve is the only way to constrain the properties of the dark matter component for local dwarfs. To this end, we have therefore searched the literature for systems with high quality rotation curve data probing the gravitational potential with good sampling, large radial extension (i.e., up to $R > R_{opt} = 3.2 R_d$ with $R_d$ the disk scale length) and small errors. Whenever possible, we rely on H$\alpha$ data or a combination of HI and H$\alpha$ so that the potential impact of beam smearing on the circular velocity is reduced and does not bias the estimate of the inner slope\footnote{A possible concern on the use of the H$\alpha$ data is related to H$\alpha$ possibly tracing biased star forming regions of the disc (as, e.g., spiral arms). As a consequence, the measured circular velocity could be non representative of the disc stars kinematics. While we can not definitively exclude such a possibility, it is nevertheless worth noting that, whenever both HI and H$\alpha$ data are available in the same region, they closely track each other. We are therefore confident that no significant bias is induced by the use of H$\alpha$ data}. The data typically provide the total circular velocity so that one needs first to model the baryons (stars and gas) contribution. While this is not a problem for the gas (whose mass profile is directly obtained from HI data), an estimate of the stellar mass\,-\,to\,-\,light ratio $\Upsilon_{\star}$ is needed in order to convert the observed luminosity profile into a mass one. Should galaxy colors be available, one could rely on stellar population synthesis models to infer $\Upsilon_{\star}$ which is indeed the case for most of the galaxies in our sample. On the contrary, when this is not the case, a fiducial value can be adopted based on the prior constraints on $\Upsilon_{\star}$. We do not revisit here this issue setting $\Upsilon_{\star}$ to the value adopted in the paper which the data are presented in. However, in order to take care of possible deviations from this value (due to possibly wrong assumptions or uncertainties in the different quantities entering stellar population synthesis codes), we add a free parameter $\kappa_{\star} = \Upsilon_{\star}/\Upsilon_{fid}$ with $\Upsilon_{\star}$ the actual unknown stellar $M/L$ ratio and $\Upsilon_{fid}$ the assumed one in the original paper. Note that $\kappa_{\star}$ can also be smaller than unity, a typical example being the assumption of an incorrect initial mass function (scaling down the $M/L$ estimates by a constant factor smaller than unity). We therefore conservatively set a flat prior on $\kappa_{\star}$ over the range $(0.55, 1.45)$ so that we fully take into account both the systematic uncertainties on the initial IMF choice and the scatter introduced by variations in metallicity and star formation rate. Although we are mainly interested in dwarfs, we initially select also LSB galaxies since they share most properties with the dwarfs ones. We finally end up with a sample of 41 galaxies reducing to 37 (28 dwarfs and 9 LSBs) after removing multiple cases. It is, indeed, sometime possible that the same galaxy has multiple measurement of the rotation curve in which case we choose the better quality one. We have, however, checked that the results fitting the other dataset are consistent thus making us confident that no bias has been induced by this choice. Table 1 gives the list of galaxies and the references for the data on the rotation curve. In more detail, dwarfs galaxies mainly come from the \cite{Sw11} and \cite{Oh} samples and have a typical dynamical mass $\sim 10^9 \ {\rm M_{\odot}}$. Another dwarf spiral (NGC 2976) comes from the THINGS sample which also gives us two dwarfs irregular (IC 2574 and NGC 2366). Finally, the sample in \cite{S05} contributes the dwarf spiral NGC 4605, the SBc galaxy NGC 5649 and the Sc one NGC 5963, and the low mass spiral NGC 6689 with different listings for its morphological type. LSBs systems mainly comes from the \cite{dBB02} sample which we refer the reader for details.

In order to fit these data, we have first to choose a density profile for the dark matter distribution. To this end, one can rely on the outcome of N\,-\,body simulations choosing a double power\,-\,law model, i.e. $\rho \propto x^{-\alpha_{in}} (1 + x)^{-\alpha_{out} + \alpha_{in}}$, with $x = r/R_s$ and $(\alpha_{in}, \alpha_{out}, R_s)$ setting the shape of the profile. While there is a large consensus on $\alpha_{out} = 3$, a still open debate exists on which $\alpha_{in}$ value better fits the numerical simulations and whether this is or not a universal quantity (see, e.g., \citealt{NFW97,Moore+98,JS00,R03,D04}). Actually, more recent simulations suggests that the logarithmic density slope $\alpha = d\ln{\rho_{DM}}/d\ln{r}$ never asymptotes a constant value in the small $r$ limit, while an exponential\,-\,like decline is achieved in the outskirts. Following \cite{N04}, we therefore assume that the halo density profile is given by the Einasto (1965, 1969, see also \citealt{PoLLS,ML05,Gr06}) model\,:

\begin{equation}
\rho_{DM}(r) = \rho_{-2} \exp{\left \{ - 2 n_{DM} \left [ \left ( \frac{r}{R_{-2}} \right )^{\frac{1}{n_{DM}}} - 1 \right ] \right \}}
\end{equation}
with $n_{DM}$ a slope parameter, $R_{-2}$ a characteristic radius defined so that $\alpha(R_{-2}) = -2$ and $\rho_{-2}$ a scaling density. It is convenient to reparameterize the model replacing the $(R_{-2}, \rho_{-2})$ parameters with $(c_{vir}, M_{vir})$ where we have introduced the concentration $c_{vir} = R_{-2}/R_{vir}$, the virial mass $M_{vir}$ (defined as the mass within a sphere containing $\Delta_{vir} \simeq 337$ times the cosmic mean matter density $\bar{\rho}_M =  3 H_0^2 \Omega_M/8 \pi G$) and the virial radius $R_{vir} = (3 M_{vir}/4 \pi \Delta_{vir} \bar{\rho}_M)^{1/3}$. Actually, rather than $M_{vir}$, we will use as a parameter $V_{vir} = (G M_{vir}/R_{vir})^{1/2}$, i.e. the circular velocity at the virial radius. Assuming spherical symmetry, the mass profile simply reads\,:

\begin{equation}
M(r) = \frac{4 \pi n_{DM} R_{-2}^3 \rho_{-2} {\rm e}^{2 n_{DM}}}{(2 n_{DM})^{3 n_{DM}}} \ \times \ \gamma\left(3 n_{DM}, \frac{r}{R_{-2}} \right )
\label{eq: mass}
\end{equation}
with $\gamma(3n_{DM}, x)$ the incomplete gamma function. The circular velocity is then $v_c^2(r) = G M(r)/r$ and may be analytically estimated as function of the the model parameters.

As already hinted at above, we parameterize the Einasto model through the three quantities $(n_{DM}, c_{vir}, V_{vir})$ to which we add $\kappa_{\star}$ to take care of the uncertainties on the stellar $M/L$ ratio. In order to determine the four quantities $(\kappa_{\star}, n_{DM}, c_{vir}, V_{vir})$, we first define the likelihood function ${\cal{L}}({\bf p}) \propto \exp{[-\chi^2({\bf p})/2]}$ with

\begin{displaymath}
\chi^2({\bf p}) = \sum_{i = 1}^{{\cal{N}}_{obs}}{\left [ \frac{v_{c,obs}{R_i} - v_{c,th}(R_i, {\bf p})}{\varepsilon_i} \right ]^2}
\end{displaymath}
where the sum is over the ${\cal{N}}_{obs}$ datapoints with $v_{c,obs}(R_i)$ and $\varepsilon_i$ the i\,-\,th measured value and its error and $v_{c,th}(R_i, {\bf p})$ the theoretically expected one. To efficiently explore the parameter space, we use a Markov Chain Monte Carlo method running three chains and checking their convergence through the Gellman--Rubin test \cite{GR92}. According to the Bayesian statistical theory, the set of parameters maximizing the likelihood do not necessarily represent the most confident constraints on each single parameter. On the contrary, one has to look at the marginalized distributions and take the median and $68\%$ confidence ranges as final constraints. To this end, we first cut out the burn\,-\,in phase of each single chain, then merge them together and use the resulting sample (after thinning to avoid spurious correlations) to estimate the median and $68\%$ confidence ranges. This sample can also be used to infer constraints on derived quantities (such as the virial mass and the logarithmic slope $\alpha$ at different radii) by estimating them along the merged chain and studying the distribution of the resulting values.

\section{Results}

\begin{figure*}
\centering
\subfigure{\includegraphics[width=5.0cm]{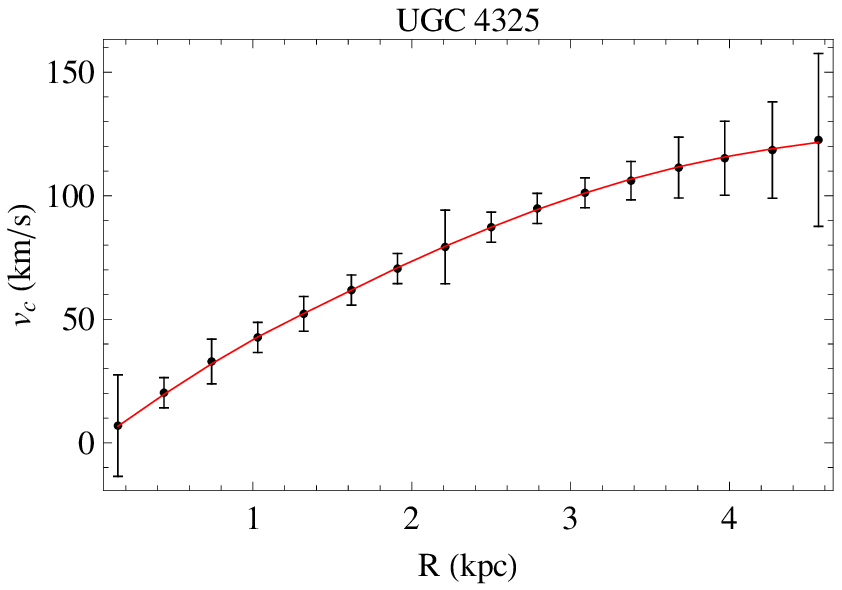}} \goodgap
\subfigure{\includegraphics[width=5.0cm]{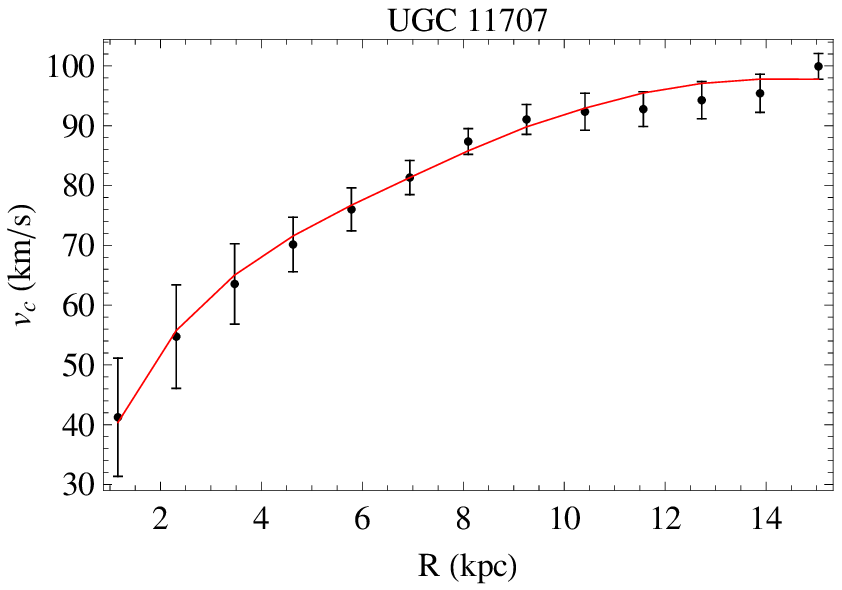}} \goodgap
\subfigure{\includegraphics[width=5.0cm]{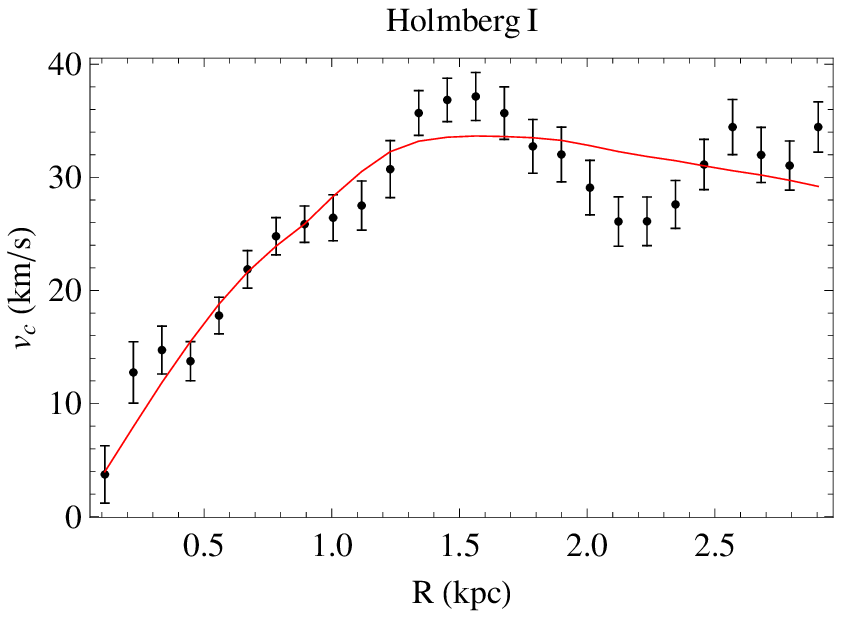}} \\
\caption{Best fit curves superimposed to the data for three typical cases with very low (left), average (centre) and high (right) $\tilde{\chi}^2$.}
\label{fig: bf}
\end{figure*}

\begin{figure*}
\centering
\subfigure{\includegraphics[width=5.0cm]{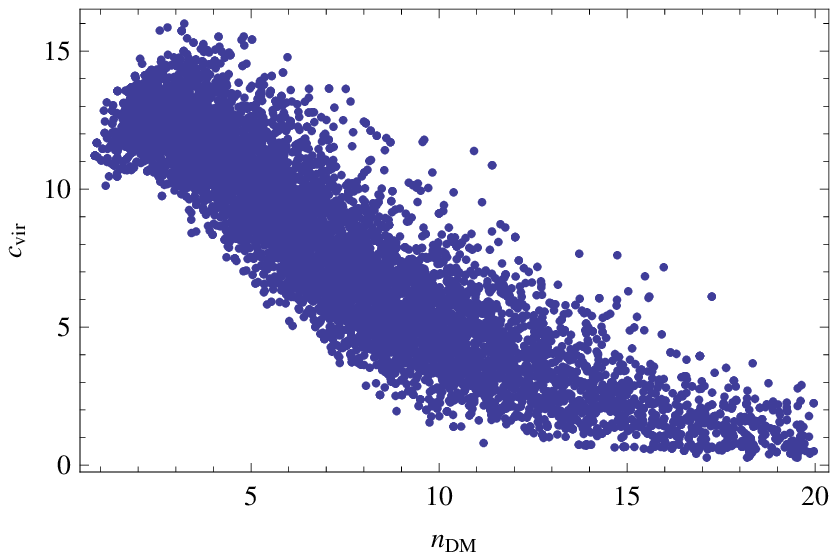}} \goodgap
\subfigure{\includegraphics[width=5.0cm]{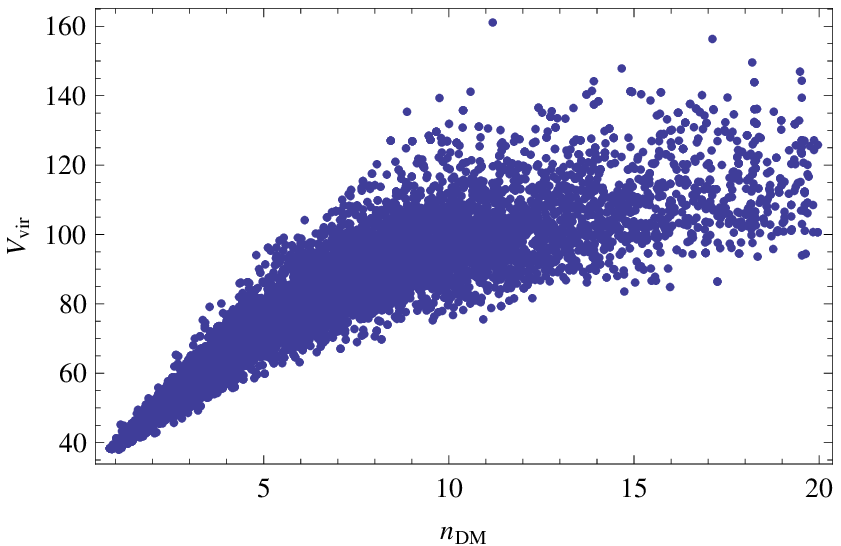}} \goodgap
\subfigure{\includegraphics[width=5.0cm]{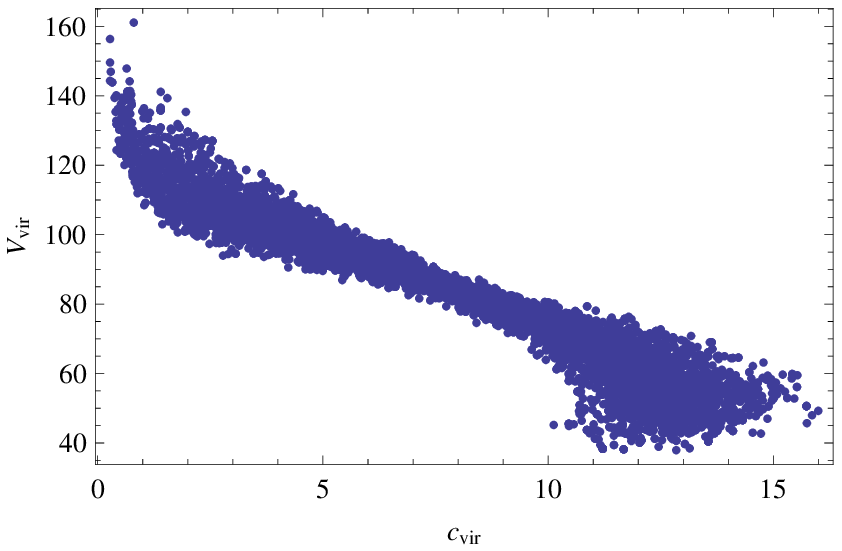}} \\
\caption{Scatter plot of the MCMC samples for the fit to the UGC 11707 data.}
\label{fig: mcmcfit}
\end{figure*}

As a general remark, we find that the Einasto model indeed fits quite well the rotation curve data for both dwarfs and LSB galaxies so that the estimated constraints on the model parameters (reported in Table 1) may be deemed as reliable\footnote{In the appendix, we will briefly present an analysis of simulated rotation curves to further check that the fitting procedure is indeed able to correctly recover the input model parameters.}. Table\,\ref{tab: fitres} also gives the values of the reduced $\chi^2$, i.e. $\tilde{\chi}^2 = \chi^2/d.o.f.$ with $d.o.f. = {\cal{N}}_{obs} - 4$ the number of degrees of freedom. Actually, this quantity should be considered only as a way to quantitatively guess the accordance of the model with the data, but one must not interpret it as the textbook $\chi^2$ estimator hence expecting $\tilde{\chi}^2 \sim 1$ for the best fit model. Indeed, the errors on the measured circular velocities are not Gaussian distributed and, moreover, typically takes into account also corrections for non circular motions and asymmetry between the approaching and receding side of the rotation curve data. Finally, for some cases (as, e.g., the de Blok \& Bosma (2002) sample), the data have been suitably smoothed so that a good model will fit the data with a very small $\tilde{\chi}^2$. This is indeed the case for many curves as can be seen, e.g., for UGC\,4325 plotted in the left panel of Fig.\,\ref{fig: bf}. While this is an overoptimistic case, the agreement within the theoretical prediction and observed rotation curve is typically quite good with the best fit curve closely interpolated the measured circular velocity (as, e.g., for UGC\,11707 shown in the central panel of Fig.\,\ref{fig: bf}). For few cases, we get a large $\tilde{\chi}^2$ which could be interpreted as a failure of the model. Actually, rather than signalling a mismatch between data and theory, a large $\tilde{\chi}^2$ is typically an evidence for the presence of wiggles in the data due to residual non circular motions or clumpy gas distribution as can be seen, for instance, in the case of Holmberg\,I in the right panel of Fig.\,\ref{fig: bf}. We therefore conclude that the Einasto model provide a good description of the DM haloes of both dwarfs and LSB galaxies and rely on it to estimate the quantities of interest for the following analysis.

As it is apparent from the $68\%$ confidence ranges reported in Table\,\ref{tab: fitres}, the constraints on the fitting parameters are not quite stringent. Needless to say, this is a consequence of both the data quality and the degeneracy in the 4D parameter space. Indeed, in order to break the degeneracies, one should not only trace the circular velocity up to large $R/R_d$ values (as is the case for most galaxies in our sample), but also reduce as much as possible the measurement uncertainties and finely sample the rotation curve. Indeed, the smaller errors are obtained for the galaxies in the THINGS \citep{Things} sample which are close to fulfill these requirements. Actually, also in these cases, the constraints on the virial velocity and hence the virial mass can be weak if the data do not probe the outer halo dominated regions. As a consequence, the stellar $M/L$ ratio becomes degenerate with $\log{M_{vir}}$ since both quantities scale the relative contribution of the stars and DM to the total $v_c(r)$. Lacking a strong determination of $\Upsilon_{\star}$, one can not weight the DM contribution in the disk dominated regions so that the uncertainties on the slope $n_{DM}$ and concentration $c_{vir}$ increase. Notwithstanding these problems, it is nevertheless interesting to look at the distribution of the halo parameters $(n_{DM}, c_{vir}, \log{M_{vir}})$ where here and in the following we have replaced the virial velocity with the logarithm of the virial mass since this latter quantity provides a better characterization of the DM halo. Moreover, we impose a loose cut on $(n_{DM}, c_{vir}, \log{M_{vir}})$ excluding galaxies with $n_{DM} > 11$ or $\log{M_{vir}} < 9.0$ since they are likely outliers thus ending up with a sample made by 27 dwarfs and 8 LSBs.

As a first test, we compare the distribution of the model parameters for dwarfs and LSBs. The mean, median and $68\%$ confidence ranges of $(n_{DM}, c_{vir}, \log{M_{vir}})$ turn to be\footnote{Hereafter, we denote with $\langle x \rangle$ ($\hat{x}$) the mean (median) $x$ value.}\,:

\begin{displaymath}
\langle n_{DM} \rangle = 3.05 \ , \ \hat{n}_{DM} = 2.58 \ , \ 68\% \ CL = (1.29, 4.36) \ ,
\end{displaymath}

\begin{displaymath}
\langle c_{vir} \rangle = 13.51 \ , \ \hat{c}_{vir} = 12.76 \ , \ 68\% \ CL = (8.14, 20.81) \ ,
\end{displaymath}

\begin{displaymath}
\langle \mu_{vir} \rangle = 10.71 \ , \ \hat{\mu}_{vir} = 10.59 \ , \ 68\% \ CL = (9.99, 11.42) \ ,
\end{displaymath}
for the dwarfs sample (with $\mu_{vir} = \log{M_{vir}}$) and\,:

\begin{displaymath}
\langle n_{DM} \rangle = 1.48 \ , \ \hat{n}_{DM} = 1.49 \ , \ 68\% \ CL = (1.00, 2.00) \ ,
\end{displaymath}

\begin{displaymath}
\langle c_{vir} \rangle = 14.13 \ , \ \hat{c}_{vir} = 11.39 \ , \ 68\% \ CL = (8.23, 18.28) \ ,
\end{displaymath}

\begin{displaymath}
\langle \mu_{vir} \rangle = 10.73 \ , \ \hat{\mu}_{vir} = 10.76 \ , \ 68\% \ CL = (10.07, 11.48) \ ,
\end{displaymath}
for the LSBs sample. Although the halo mass range probed is almost the same, the slope parameter $n_{DM}$ takes larger values for the dwarfs than for the LSBs. This is essentially due to the strongly asymmetric $n_{DM}$ distribution for the dwarfs having a long tail towards high $n_{DM}$. As a consequence, both the mean and the median are larger than for the LSBs. On the contrary, we find no significative difference for the concentration parameter with the $68\%$ confidence range being almost perfectly overlapped. Actually, the statistical significance of the $n_{DM}$ difference is undermined by the large errors on the individual determinations. Moreover, the statistical properties of the halo parameters distribution for the joint sample, namely

\begin{displaymath}
\langle n_{DM} \rangle = 2.69 \ , \ \hat{n}_{DM} = 2.22 \ , \ 68\% \ CL = (1.17, 4.08) \ ,
\end{displaymath}

\begin{displaymath}
\langle c_{vir} \rangle = 13.65 \ , \ \hat{c}_{vir} = 12.76 \ , \ 68\% \ CL = (8.14, 20.81) \ ,
\end{displaymath}

\begin{displaymath}
\langle \mu_{vir} \rangle = 10.71 \ , \ \hat{\mu}_{vir} = 10.60 \ , \ 68\% \ CL = (9.99, 11.48) \ ,
\end{displaymath}
are quite similar to the dwarfs one so that we prefer to increase the statistics adding LSBs to the dwarfs sample. Although such a strategy decreases the homogeneity of the sample, we have checked that it does not bias the results of interest as could be anticipated noting that the final sample is still dominated by dwarfs. In the following, for sake of simplicity, we will refer to the full set as the dwarfs sample.

\begin{figure}
\centering
\includegraphics[width=7.5cm]{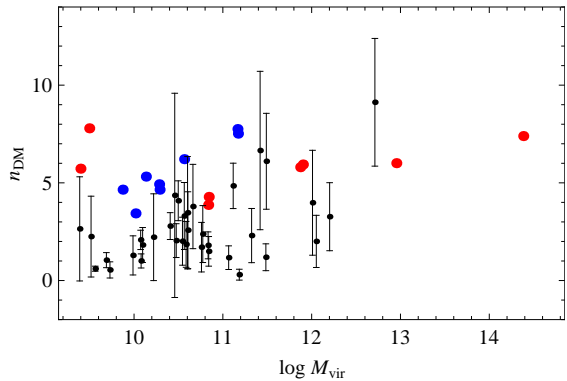}
\caption{Einasto slope parameter $n_{DM}$ vs the virial mass $\log{M_{vir}}$ for our sample (black) and the N\,-\,body haloes of Navarro et al. (2004, blue) and Merritt et al. (2006, red).}
\label{fig: ndmmvirplot}
\end{figure}

Since N\,-\,body simulations point at the Einasto model as the best representation of the dark haloes density profile, the successful fits of the dwarfs galaxies rotation curves may be seen as a further evidence of this model virtues. Actually, our result tells us something more since it refers to a mass range typically not resolved by simulations. It is therefore interesting to compare $n_{DM}$ distribution we find with the ones predicted by the numerical simulations. These works find values for $n_{DM}$ covering the range $(5.0, 7.0)$ in a roughly uniform way \citep{N04,Mer06,Gr06,N10}, while our $n_{DM}$ distribution, although overlapping with the predicted one, is definitely more populated by shallower profiles. This can also be read from Fig.\,\ref{fig: ndmmvirplot} where we plot $n_{DM}$ vs $\log{M_{vir}}$ for both our data and the results from N\,-\,body simulations \citep{N04,Mer06}. A straightforward comparison is, however, not possible given both the different mass range and the uncertain impact of baryons (which are not accounted for in the simulations and may change the density slope). In order to partially alleviate this problem, it is much more instructive to compare our $n_{DM}$ values to those obtained by fitting the full THINGS sample. \cite{CdBM11} have indeed found that the Einasto model fits the observed rotation curves in a very good way being also superior to popular models as the NFW and the isothermal ones. Moreover, they find that better fits are obtained by adopting a Kroupa (2001) IMF rather than the fiducial diet Salpeter used by the THINGS team. This is the same as downscaling the stellar $M/L$ by a factor $\sim 1/1.16 \simeq 0.86$ which is consistent with the typical values we find for our $\kappa_{\star}$ parameter. \cite{CdBM11} have also found that their halo sample can be divided in four classes according to the $n_{DM}$ value with the most numerous class being represented by systems with $0.1 \le n_{DM} \le 4.0$ and the rest of galaxies mainly having $n_{DM} > 4.0$ values. This is roughly consistent with what we find here thus suggesting that dwarfs and normal spiral galaxies cover the same range for the slope of the Einasto model. However, since our confidence ranges are quite large and strong degeneracies with the other model parameters are present (as can be seen from the distribution of the MCMC points in Fig.\,\ref{fig: mcmcfit}), we prefer to not speculate further on this issue, but only point out at this qualitative agreement as a further evidence of the reliability of our choice of the Einasto model for the dwarfs galaxies haloes.

\section{Dark halo properties}

\begin{table*}
\scriptsize
\caption{Correlation properties of the Einasto model parameters with the environment. Rows are as follows\,: 1. Spearman correlation coefficient; 2., 3., 4. median and $68\%$ CL of $(a, b, \sigma_{int})$. The number of galaxies used is ${\cal{N}} = 13, 29, 33$ for correlations with $\Theta$, $\Sigma_{10}$, $D_{750}$, respectively. See text for details on the fitted quantities and the median values of the input parameters.}
\begin{center}
\begin{tabular}{cccccccccc}
\hline
Id & $n_{DM}$\,-\,$\Theta$ & $n_{DM}$\,-\,$\Sigma_{10}$ & $n_{DM}$\,-\,$D_{750}$ & $c_{vir}$\,-\,$\Theta$ & $c_{vir}$\,-\,$\Sigma_{10}$ & $c_{vir}$\,-\,$D_{750}$ & $\mu_{vir}$\,-\,$\Theta$ & $\mu_{vir}$\,-\,$\Sigma_{10}$ & $\mu_{vir}$\,-\,$D_{750}$ \\
\hline
${\cal{C}}(x, y)$ & $-0.16 \pm 0.12$ & $-0.07 \pm 0.04$ & $0.29 \pm 0.03$ & $-0.63 \pm 0.08$ & $-0.11 \pm 0.04$ & $-0.16 \pm 0.02$ & $0.36 \pm 0.09$ & $-0.20 \pm 0.04$ & $0.32 \pm 0.03$ \\

~  & ~  & ~  & ~  & ~  & ~  & ~  & ~  & ~  & ~  \\

$a$ & $-0.04_{-0.05}^{+0.06}$ & $0.00_{-0.02}^{+0.02}$ & $0.07_{-0.10}^{+0.21}$ & $-0.07_{-0.05}^{+0.07}$ & $0.00_{-0.01}^{+0.01}$ & $-0.11_{-0.14}^{+0.12}$ & $0.07_{-0.08}^{+0.11}$ & $-1.09_{-1.00}^{+0.58}$ & $0.72_{-0.59}^{+0.68}$ \\

%$a$ & $-0.036_{-0.053}^{+0.058}$ & $-0.003_{-0.019}^{+0.021}$ & $0.072_{-0.095}^{+0.205}$ & $-0.067_{-0.048}^{+0.065}$ & $-0.002_{-0.004}^{+0.005}$ & $-0.105_{-0.141}^{+0.119}$ & $0.074_{-0.077}^{+0.114}$ & $-1.090_{-1.003}^{+0.575}$ & $0.718_{-0.586}^{+0.676}$ \\

~  & ~  & ~  & ~  & ~  & ~  & ~  & ~  & ~  & ~  \\

$b$ & $-0.19_{-0.01}^{+0.01}$ & $-0.04_{-0.02}^{+0.02}$ & $-0.13_{-0.45}^{+0.26}$ & $0.09_{-0.01}^{+0.01}$ & $0.04_{-0.01}^{+0.01}$ & $0.18_{-0.13}^{+0.17}$ & $-0.50_{-0.04}^{+0.06}$ & $1.14_{-0.75}^{+0.74}$ & $-0.96_{-0.81}^{+0.66}$ \\

%$b$ & $-0.187_{-0.012}^{+0.012}$ & $-0.040_{-0.024}^{+0.022}$ & $-0.131_{-0.449}^{+0.263}$ & $0.093_{-0.005}^{+0.002}$ & $0.039_{-0.013}^{+0.009}$ & $0.177_{-0.128}^{+0.167}$ & $-0.499_{-0.036}^{+0.057}$ & $1.143_{-0.749}^{+0.736}$ & $-0.962_{-0.806}^{+0.660}$ \\

~  & ~  & ~  & ~  & ~  & ~  & ~  & ~  & ~  & ~  \\

$\sigma_{int}$ & $0.22_{-0.06}^{+0.07}$ & $0.23_{-0.04}^{+0.06}$ & $0.22_{-0.04}^{+0.05}$ & $0.15_{-0.03}^{+0.04}$ & $0.17_{-0.03}^{+0.03}$ & $0.16_{-0.02}^{+0.03}$ & $0.28_{-0.08}^{+0.11}$ & $0.37_{-0.16}^{+0.12}$ & $0.46_{-0.09}^{+0.12}$ \\

%$\sigma_{int}$ & $0.224_{-0.055}^{+0.071}$ & $0.230_{-0.042}^{+0.055}$ & $0.225_{-0.040}^{+0.046}$ & $0.154_{-0.031}^{+0.043}$ & $0.167_{-0.025}^{+0.030}$ & $0.159_{-0.023}^{+0.028}$ & $0.277_{-0.082}^{+0.106}$ & $0.367_{-0.155}^{+0.119}$ & $0.460_{-0.091}^{+0.118}$ \\
\hline
\end{tabular}
\end{center}
\label{tab: corrcoeffhalopar}
\end{table*}

Having convincingly demonstrated that the Einasto model provides a good fit to the rotation curve data, we can use the constraints derived on its parameters to estimate other quantities of interest. To this end, one has only to evaluate $y = y(n_{DM}, c_{vir}, \log{M_{vir}})$ along the final merged chain and then look at the distribution of the $y$ values. It is then instructive to look for correlations of the estimated quantities with both the stellar luminosity and size and the environment the dwarf resides in. To this regard, it is worth stressing that our aim here is not to constrain any particular galaxy formation scenario, but mainly to search for any evidence of scaling relation which could then be used as a guidance to improve our understanding of haloes assembly.

\subsection{Preliminary remarks}

Before discussing the results of our analysis, we have first to explain what are the estimator we use to describe the environment a galaxy lives in. To this end, we use three different statistics to quantify whether a galaxy lives in a dense environment or is rather an isolated one. First, we consider the {\it tidal index} \citep{KM96}

\begin{equation}
\Theta = {\rm max}[\log{(M_k/D_k^3}] + C \ ,
\label{eq: defti}
\end{equation}
where $(M_k, D_k)$ are the mass and distance to the chosen galaxy of the $k$\,-\,th system and $C$ is a normalization constant. Estimating $\Theta$ for a given galaxy is actually quite difficult and typically possible only for systems in the Local Group. A cross match between our sample and the Catalog of Neighboring Galaxies \citep{CNG04} allows us to infer $\Theta$ for 10 dwarfs and 4 LSBs only.

In order to further test a possible correlation between the inner slope and the environment, we resort to two other commonly used indicators. The first, denoted as $\Sigma_{10}$, is the projected number density of galaxies inside the circle of radius equal to the projected distance of the tenth nearest neighbor \citep{D80}, while the second, denoted as $D_{750}$, is the projected number density of galaxies within a circle of radius equal to 750 kpc centred on the system of interest \citep{T09}. We will refer to $\Sigma_{10}$ and $D_{750}$ as the {\it local} and {\it global} overdensity and stress that both these quantities are normalized with respect to their mean value over the sample so that galaxies with $\Sigma_{10}$ and $D_{750}$ larger (smaller) than unity are in over (under) dense regions. As a general remark, we note that, although easy to calculate, both $\Sigma_{10}$ and $D_{750}$ are looser indicators of dynamical state of a galaxy with respect to the tidal index. For instance, a galaxy could have a large value of $\Sigma_{10}$, but a small $\Theta$ if it is embedded in an environment rich of small mass systems. In such a case, the galaxy should be considered isolated even if $\Sigma_{10}$ is large. However, such case should be rare and, indeed, we find that both $\Sigma_{10}$ and $D_{750}$ positively correlate with $\Theta$ so that we use them as supplementary tools to check whether the environment impacts the halo density profile. Using the NED database\footnote{{\tt nedwww.ipac.caltech.edu}}, we are able to estimate $\Sigma_{10}$ for 30 galaxies (22 dwarfs and 8 LSB), while $D_{750}$ is available for 34 systems (26 dwarfs and 8 LSBs) thus doubling the sample usable for the correlations with the tidal index.

\begin{figure*}
\centering
\subfigure{\includegraphics[width=5.0cm]{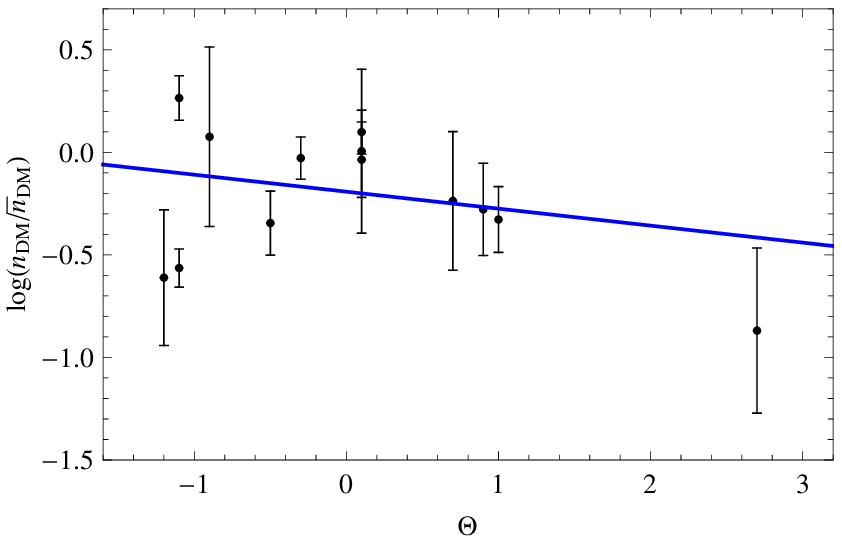}} \goodgap
\subfigure{\includegraphics[width=5.0cm]{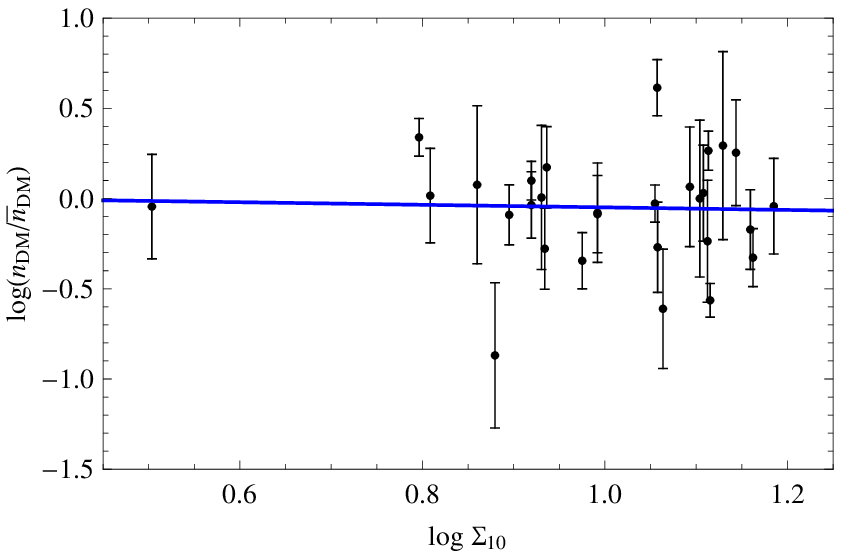}} \goodgap
\subfigure{\includegraphics[width=5.0cm]{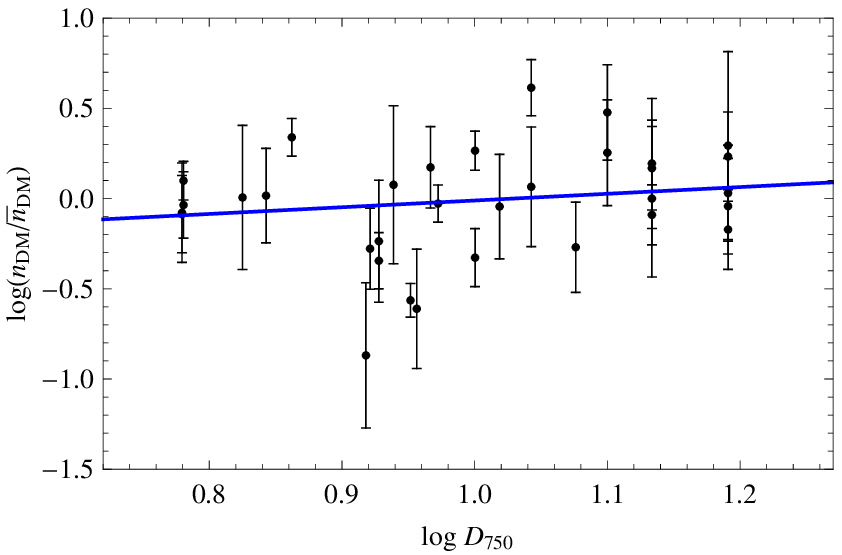}} \\
\caption{Median normalized Einasto slope parameter $n_{DM}$ vs the three environment indicators.}
\label{fig: ndmplot}
\end{figure*}

\begin{table*}
%\scriptsize
\caption{Same as Table\,\ref{tab: corrcoeffhalopar} but for the logarithmic slopes $(\alpha_d, \alpha_{0.1})$ using ${\cal{N}} = 12, 25, 28$ galaxies for correlations with $(\Theta$, $\Sigma_{10}$, $D_{750})$.}
\begin{center}
\begin{tabular}{ccccccc}
\hline
Id & $\alpha_{d}$\,-\,$\Theta$ & $\alpha_{d}$\,-\,$\Sigma_{10}$ & $\alpha_{d}$\,-\,$D_{750}$ & $\alpha_{0.1}$\,-\,$\Theta$ & $\alpha_{0.1}$\,-\,$\Sigma_{10}$ & $\alpha_{0.1}$\,-\,$D_{750}$ \\
\hline
${\cal{C}}(x, y)$ & $-0.27 \pm 0.10$ & $-0.14 \pm 0.05$ & $0.24 \pm 0.04$ & $-0.40 \pm 0.09$ & $-0.07 \pm 0.04$ & $-0.19 \pm 0.04$ \\

~  & ~  & ~  & ~  & ~  & ~  & ~ \\

$a$ & $0.00_{-0.03}^{+0.02}$ & $0.00_{-0.01}^{+0.01}$ & $0.21_{-0.16}^{+0.30}$ & $-0.04_{-0.04}^{+0.04}$ & $0.01_{-0.04}^{+0.03}$ & $-0.03_{-0.15}^{+0.12}$ \\

~  & ~  & ~  & ~  & ~  & ~  & ~ \\

$b$ & $0.96_{-0.01}^{+0.02}$ & $0.99_{-0.02}^{+0.01}$ & $0.77_{-0.36}^{+0.29}$ & $1.02_{-0.03}^{+0.01}$ & $0.92_{-0.04}^{+0.04}$ & $0.99_{-0.13}^{+0.23}$ \\

~  & ~  & ~  & ~  & ~  & ~  & ~ \\

$\sigma_{int}$ & $0.38_{-0.08}^{+0.11}$ & $0.36_{-0.05}^{+0.07}$ & $0.38_{-0.06}^{+0.06}$ & $0.20_{-0.05}^{+0.08}$ & $0.16_{-0.03}^{+0.05}$ & $0.15_{-0.02}^{+0.03}$ \\
\hline
\end{tabular}
\end{center}
\label{tab: corrcoeffslope}
\end{table*}

To quantify whether there exists any correlation between a given quantity $y(n_{DM}, c_{vir}, \log{M_{vir}})$ correlates and the environment indicators $(\Theta, \Sigma_{10}, D_{750})$, we use two different approaches. First, in order to get a quick result, we compute the Spearman rank correlation coefficient denoted as ${\cal{C}}(x, y)$. Since the sample used is not large enough, it is possible that few points drive the ${\cal{C}}(x, y)$ estimate. In order to avoid such a bias, we resort to a jacknife approach evaluating ${\cal{C}}(x, y)$ for ${\cal{N}} - 1$ samples obtained by excluding one galaxy at time and use the mean value and the standard deviation of this sample as our final estimate. To quantify whether the result indicates the presence or lack of a statistically significant correlation, we can compute the $z$\,-\,score defined as $z = [({\cal{N}} - 3)/1.06]^{1/2} {\rm arctanh}[{{\cal{C}}(x, y)}]$ where ${\cal{N}}$ is the number of datapoints. Since $z$ approximately follows a standard normal distribution, one can quickly check whether the ${\cal{C}}(x, y)$ value corresponds to a statistically meaningful correlation.

\begin{figure*}
\centering
\subfigure{\includegraphics[width=5.0cm]{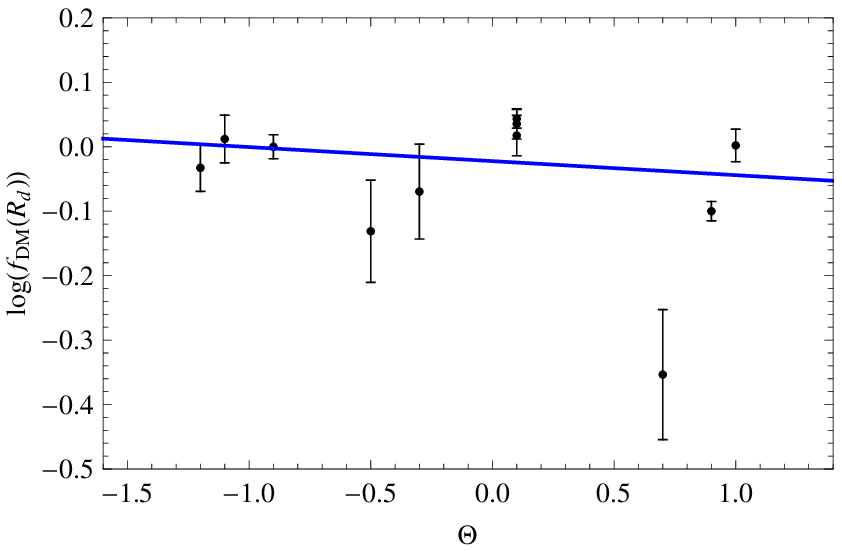}} \goodgap
\subfigure{\includegraphics[width=5.0cm]{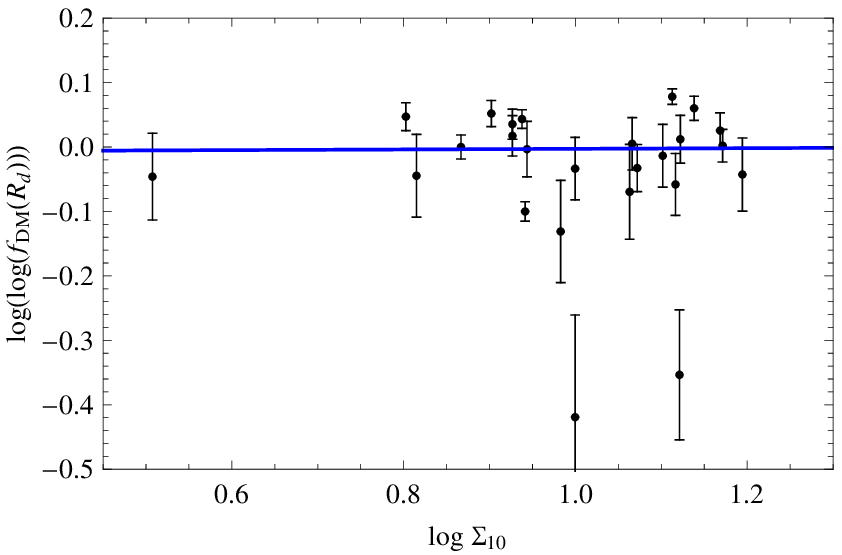}} \goodgap
\subfigure{\includegraphics[width=5.0cm]{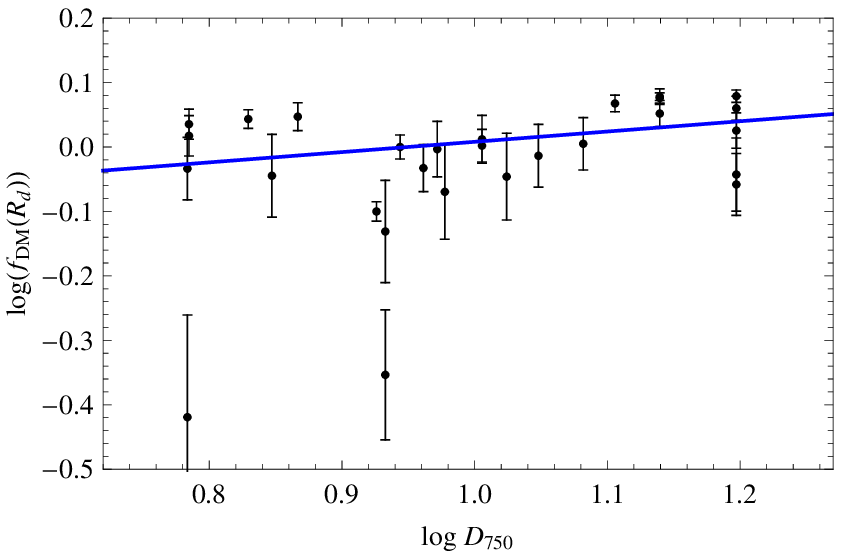}} \\
\caption{Same as Fig.\,\ref{fig: ndmplot} but for the DM mass fraction at the disc scalelength.}
\label{fig: fdmplot}
\end{figure*}

The Spearman rank estimator does not take into account the errors on $y$ coming out from the uncertainties on the constrained halo parameters. We therefore rely on a different approach briefly sketched below. Let us suppose that a linear relation exists among two different quantities denoted $(R, Q)$ such as $R = a  Q + b$ and denote with $\sigma_{int}$ the intrinsic scatter around this relation. Calibrating such a relation means determining the two coefficients $(a, b)$ and the intrinsic scatter $\sigma_{int}$. To this aim, we will resort to a Bayesian motivated technique \cite{Dago05} thus maximizing the likelihood function ${\cal{L}}(a, b, \sigma_{int}) = \exp{[-L(a, b, \sigma_{int})]}$ with\,:

\begin{eqnarray}
L(a, b, \sigma_{int}) & = & \frac{1}{2} \sum{\ln{(\sigma_{int}^2 + \sigma_{R_i}^2 + a^2 \sigma_{Q_i}^2)}} \nonumber \\
~ & + &
\frac{1}{2} \sum{\frac{(R_i - a Q_i - b)^2}{\sigma_{int}^2 + \sigma_{Q_i}^2 + a^2 \sigma_{Q_i}^2}}
\label{eq: deflike}
\end{eqnarray}
where the sum is over the ${\cal{N}}$ objects in the sample. Note that, actually, this maximization is performed in the two parameter space $(a,
\sigma_{int})$ since $b$ may be estimated analytically as\,:

\begin{displaymath}
b = \left [ \sum{\frac{R_i - a Q_i}{\sigma_{int}^2 + \sigma_{R_i}^2 + a^2
\sigma_{Q_i}^2}} \right ] \left [\sum{\frac{1}{\sigma_{int}^2 + \sigma_{R_i}^2 + a^2
\sigma_{Q_i}^2}} \right ]^{-1}
\end{displaymath}
so that we will not consider it anymore as a fit parameter.

Two caveats are in order here. First, as we will see later, we are actually looking for power\,-\,law like relations so that, when needed, we will use logarithmic units in order to linearize them. As a second issue, we note that Eq.(\ref{eq: deflike}) implicitly assumes that the error on the quantities involved are equals on both the positive and negative side, i.e., the $68\%$ confidence range is symmetric around the mean. This is not the case for most of the quantities we are interested in so that, following \cite{Dago04}, we take the mean of the positive and negative errors as input to the likelihood function. Moreover, we typically fit $\log{(y/\hat{y})}$ rather than $\log{y}$ in order to reduce the degeneracy among the slope $a$ and intrinsic scatter $\sigma_{int}$. We finally look at the median and $68\%$ confidence range of the slope $a$ to check whether the two quantities are correlated or not noting that a null value is a definitive evidence for a lack of any correlation.

\subsection{Halo model parameters}

We start investigating correlations between the environment and the halo model parameters $(n_{DM}, c_{vir}, \mu_{vir})$ where, hereafter, we will use the shortened notation $\mu_{vir}$ to refer to $\log{M_{vir}}$. The results obtained are summarized in Table\,\ref{tab: corrcoeffhalopar}.

Let us first consider the correlations involving the Einasto slope parameter $n_{DM}$ which we fit as a function\footnote{Here and in the rest of the paper, we will use logarithmic variables for $(\Sigma_{10}, D_{750})$ so that correlations with this parameter will be of the form $y \propto x^{a}$ with $y$ the quantity of interest and $x = (\Sigma_{10}, D_{750})$. On the contrary, we will retain linear units for $\Theta$ since the tidal index is already a logarithmic quantity.} of $(\Theta, \log{\Sigma_{10}}, \log{D_{750}})$ using as variable $\log{(n_{DM}/\hat{n}_{DM})}$ with $\hat{n}_{DM} = 2.22$ the median value. Both the Spearman correlation coefficient ${\cal{C}}$ and the slope $a$ of the fitted relations strongly indicate that $n_{DM}$ does not correlate with the environment. Actually, although we refer to it as the slope parameter of the Einasto profile, the logarithmic density is not related to $n_{DM}$ only, but rather to the full set of halo parameters being

\begin{table*}
%\scriptsize
\caption{Same as Table\,\ref{tab: corrcoeffhalopar} but for the DM mass content $(f_d, f_{-2})$ using ${\cal{N}} = 11, 25, 28$ galaxies for correlations with $(\Theta$, $\Sigma_{10}$, $D_{750})$.}
\begin{center}
\begin{tabular}{ccccccc}
\hline
Id & $f_{d}$\,-\,$\Theta$ & $f_{d}$\,-\,$\Sigma_{10}$ & $f_{d}$\,-\,$D_{750}$ & $f_{-2}$\,-\,$\Theta$ & $f_{-2}$\,-\,$\Sigma_{10}$ & $f_{-2}$\,-\,$D_{750}$ \\
\hline
${\cal{C}}(x, y)$ & $-0.03 \pm 0.11$ & $-0.02 \pm 0.04$ & $0.39 \pm 0.04$ & $-0.29 \pm 0.16$ & $-0.17 \pm 0.05$ & $0.06 \pm 0.04$ \\

~  & ~  & ~  & ~  & ~  & ~  & ~ \\

$a$ & $-0.01_{-0.01}^{+0.02}$ & $0.00_{-0.02}^{+0.03}$ & $0.13_{-0.06}^{+0.06}$ & $-0.002_{-0.006}^{+0.005}$ & $-0.04_{-0.05}^{+0.03}$ & $0.001_{-0.003}^{+0.003}$ \\

~  & ~  & ~  & ~  & ~  & ~  & ~ \\

$b$ & $-0.02_{-0.01}^{+0.01}$ & $0.01_{-0.03}^{+0.01}$ & $-0.13_{-0.18}^{+0.06}$ & $-0.006_{-0.003}^{+0.008}$ & $0.06_{-0.03}^{+0.03}$ & $0.01_{-0.01}^{+0.08}$ \\

~  & ~  & ~  & ~  & ~  & ~  & ~ \\

$\sigma_{int}$ & $0.05_{-0.02}^{+0.02}$ & $0.05_{-0.01}^{+0.01}$ & $0.04_{-0.01}^{+0.01}$ & $0.003_{-0.002}^{+0.004}$ & $0.02_{-0.01}^{+0.01}$ & $0.02_{-0.01}^{+0.01}$ \\
\hline
\end{tabular}
\end{center}
\label{tab: corrcoefffd}
\end{table*}

\begin{table*}
%\scriptsize
\caption{Same as Table\,\ref{tab: corrcoeffhalopar} but for the size ratios $({\cal{R}}_{-2}, {\cal{R}}_{vir})$ using ${\cal{N}} = 11, 25, 28$ galaxies for correlations with $(\Theta$, $\Sigma_{10}$, $D_{750})$.}
\begin{center}
\begin{tabular}{ccccccc}
\hline
Id & ${\cal{R}}_{-2}$\,-\,$\Theta$ & ${\cal{R}}_{-2}$\,-\,$\Sigma_{10}$ & ${\cal{R}}_{-2}$\,-\,$D_{750}$ & ${\cal{R}}_{vir}$\,-\,$\Theta$ & ${\cal{R}}_{vir}$\,-\,$\Sigma_{10}$ & ${\cal{R}}_{vir}$\,-\,$D_{750}$ \\
\hline
${\cal{C}}(x, y)$ & $0.00 \pm 0.14$ & $-0.13 \pm 0.05$ & $-0.21 \pm 0.04$ & $-0.34 \pm 0.14$ & $-0.25 \pm 0.05$ & $-0.30 \pm 0.03$ \\

~  & ~  & ~  & ~  & ~  & ~  & ~ \\

$a$ & $-0.03_{-0.06}^{+0.04}$ & $-0.06_{-0.21}^{+0.25}$ & $-0.04_{-0.15}^{+0.15}$ & $-0.03_{-0.05}^{+0.04}$ & $-0.14_{-0.24}^{+0.28}$ & $-0.15_{-0.25}^{+0.25}$ \\

~  & ~  & ~  & ~  & ~  & ~  & ~ \\

$b$ & $-0.09_{-0.01}^{+0.03}$ & $-0.02_{-0.26}^{+0.36}$ & $0.00_{-0.17}^{+0.25}$ & $-0.03_{-0.01}^{+0.01}$ & $0.21_{-0.10}^{+0.35}$ & $0.19_{-0.29}^{+0.31}$ \\

~  & ~  & ~  & ~  & ~  & ~  & ~ \\

$\sigma_{int}$ & $0.06_{-0.04}^{+0.07}$ & $0.16_{-0.03}^{+0.05}$ & $0.16_{-0.04}^{+0.04}$ & $0.24_{-0.06}^{+0.07}$ & $0.24_{-0.04}^{+0.05}$ & $0.24_{-0.04}^{+0.05}$ \\
\hline
\end{tabular}
\end{center}
\label{tab: corrcoeffrd}
\end{table*}

\begin{equation}
\alpha(r) = -2 \left ( \frac{r}{R_{-2}} \right )^{\frac{1}{n_{DM}}} = -2 \left ( \frac{c_{vir} r}{R_{vir}} \right )^{\frac{1}{n_{DM}}} \ .
\label{eq: alphaein}
\end{equation}
The slope of the density profile better characterized by $\alpha(r_s)$ where $r_s$ is some reference radius. As first possible choice, we set $r_s = R_d$ since the region around the disk scalelength is typically the best sampled by the rotation curve data. However, since $R_d$ is different from one case to another, we also use $r_s = 0.1 R_{vir}$ so that we sample $\alpha$ at the same scaled radius for all the galaxies. Moreover, as can be easily understood from Eq.(\ref{eq: alphaein}), the smaller is $r_s/R_{vir}$, the better is $\alpha(r_s)$ constrained so that setting $r_s/R_{vir} = 0.1$ represents a good compromise between reducing the uncertainties and avoiding to infer constraints on a quantity defined over a range not probed by the data. In the following, we will refer to $\alpha(R_d)$ and $\alpha(0.1 R_{vir})$ as $\alpha_d$ and $\alpha_{0.1}$, respectively. Note that $\alpha_d$ is evaluated at a smaller radius than $\alpha_{0.1}$ given that, in order to have $0.1 R_{vir}/R_d > 1$, it is sufficient that $R_{vir}/R_d > 10$ holds which is always the case.

\begin{table*}
%\scriptsize
\caption{Same as Table\,\ref{tab: corrcoeffhalopar} but for the Newtonian accelerations at $R_d$ using ${\cal{N}} = 12, 26, 29$ galaxies for correlations with $(\Theta$, $\Sigma_{10}$, $D_{750})$.}
\begin{center}
\begin{tabular}{ccccccc}
\hline
Id & $\gamma_{d}^{DM}$\,-\,$\Theta$ & $\gamma_{d}^{DM}$\,-\,$\Sigma_{10}$ & $\gamma_{d}^{DM}$\,-\,$D_{750}$ & $\gamma_{d}^{bar}$\,-\,$\Theta$ & $\gamma_{d}^{bar}$\,-\,$\Sigma_{10}$ & $\gamma_{d}^{bar}$\,-\,$D_{750}$ \\
\hline
${\cal{C}}(x, y)$ & $-0.23 \pm 0.11$ & $-0.26 \pm 0.04$ & $0.07 \pm 0.04$ & $-0.63 \pm 0.06$ & $-0.24 \pm 0.04$ & $-0.37 \pm 0.03$ \\

~  & ~  & ~  & ~  & ~  & ~  & ~ \\

$a$ & $-0.01_{-0.04}^{+0.04}$ & $-0.03_{-0.27}^{+0.23}$ & $0.000_{-0.004}^{+0.004}$ & $-0.13_{-0.10}^{+0.06}$ & $-0.05_{-0.12}^{+0.15}$ & $-0.54_{-0.46}^{+0.33}$ \\

~  & ~  & ~  & ~  & ~  & ~  & ~ \\

$b$ & $-0.08_{-0.01}^{+0.01}$ & $0.06_{-0.29}^{+0.36}$ & $-0.053_{-0.053}^{+0.006}$ & $0.07_{-0.01}^{+0.02}$ & $0.13_{-0.27}^{+0.20}$ & $0.70_{-0.44}^{+0.29}$ \\

~  & ~  & ~  & ~  & ~  & ~  & ~ \\

$\sigma_{int}$ & $0.40_{-0.08}^{+0.10}$ & $0.36_{-0.05}^{+0.06}$ & $0.33_{-0.04}^{+0.05}$ & $0.30_{-0.06}^{+0.09}$ & $0.33_{-0.04}^{+0.05}$ & $0.32_{-0.05}^{+0.04}$ \\
\hline
\end{tabular}
\end{center}
\label{tab: corrcoeffgd}
\end{table*}

\begin{table*}
%\scriptsize
\caption{Same as Table\,\ref{tab: corrcoeffhalopar} but for the Newtonian accelerations at $R_{-2}$ using ${\cal{N}} = 12, 26, 29$ galaxies for correlations with $(\Theta$, $\Sigma_{10}$, $D_{750})$.}
\begin{center}
\begin{tabular}{ccccccc}
\hline
Id & $\gamma_{-2}^{DM}$\,-\,$\Theta$ & $\gamma_{-2}^{DM}$\,-\,$\Sigma_{10}$ & $\gamma_{-2}^{DM}$\,-\,$D_{750}$ & $\gamma_{-2}^{bar}$\,-\,$\Theta$ & $\gamma_{-2}^{bar}$\,-\,$\Sigma_{10}$ & $\gamma_{-2}^{bar}$\,-\,$D_{750}$ \\
\hline
${\cal{C}}(x, y)$ & $-0.27 \pm 0.11$ & $-0.34 \pm 0.04$ & $-0.09 \pm 0.03$ & $-0.08 \pm 0.11$ & $-0.11 \pm 0.04$ & $-0.20 \pm 0.03$ \\

~  & ~  & ~  & ~  & ~  & ~  & ~ \\

$a$ & $0.00_{-0.02}^{+0.01}$ & $-0.15_{-0.21}^{+0.19}$ & $-0.02_{-0.11}^{+0.12}$ & $0.00_{-0.02}^{+0.01}$ & $-0.04_{-0.17}^{+0.17}$ & $-0.15_{-0.20}^{+0.24}$ \\

~  & ~  & ~  & ~  & ~  & ~  & ~ \\

$b$ & $0.072_{-0.004}^{+0.006}$ & $0.24_{-0.22}^{+0.25}$ & $0.09_{-0.16}^{+0.19}$ & $0.18_{-0.02}^{+0.01}$ & $-0.04_{-0.46}^{+0.20}$ & $0.25_{-0.26}^{+0.25}$ \\

~  & ~  & ~  & ~  & ~  & ~  & ~ \\

$\sigma_{int}$ & $0.26_{-0.06}^{+0.07}$ & $0.25_{-0.04}^{+0.05}$ & $0.25_{-0.04}^{+0.04}$ & $0.08_{-0.05}^{+0.05}$ & $0.27_{-0.04}^{+0.05}$ & $0.10_{-0.03}^{+0.04}$ \\
\hline
\end{tabular}
\end{center}
\label{tab: corrcoeffgs}
\end{table*}

\begin{table*}
%\scriptsize
\caption{Same as Table\,\ref{tab: corrcoeffhalopar} but for the Newtonian accelerations as a function of $({\cal{M}}_B, \log{R_d})$.}
\begin{center}
\begin{tabular}{ccccccccc}
\hline
Id & $\gamma_{d}^{DM}$\,-\,${\cal{M}}_B$ & $\gamma_{d}^{DM}$\,-\,$R_d$ & $\gamma_{d}^{bar}$\,-\,${\cal{M}}_B$ & $\gamma_{d}^{bar}$\,-\,$R_d$ & $\gamma_{-2}^{DM}$\,-\,${\cal{M}}_B$ & $\gamma_{-2}^{DM}$\,-\,$R_d$ & $\gamma_{-2}^{bar}$\,-\,${\cal{M}}_B$ & $\gamma_{-2}^{bar}$\,-\,$R_d$ \\
\hline
${\cal{C}}(x, y)$ & $-0.09 \pm 0.04$ & $-0.19 \pm 0.04$ & $0.00 \pm 0.04$ & $-0.53 \pm 0.03$ & $-0.08 \pm 0.04$ & $-0.27 \pm 0.03$ & $-0.39 \pm 0.03$ & $-0.13 \pm 0.04$ \\

~  & ~  & ~  & ~  & ~  & ~  & ~  & ~  & ~  \\

$a$ & $-0.02_{-0.03}^{+0.02}$ & $-0.09_{-0.26}^{+0.13}$ & $0.000_{-0.004}^{+0.004}$ & $-0.66_{-0.19}^{+0.21}$ & $-0.01_{-0.02}^{+0.02}$ & $-0.24_{-0.19}^{+0.15}$ & $-0.04_{-0.03}^{+0.04}$ & $-0.02_{-0.07}^{+0.07}$ \\

~  & ~  & ~  & ~  & ~  & ~  & ~  & ~  & ~  \\

$b$ & $-0.55_{-0.50}^{+0.53}$ & $-0.04_{-0.03}^{+0.03}$ & $-0.08_{-0.11}^{+0.08}$ & $0.14_{-0.04}^{+0.02}$ & $-0.29_{-0.62}^{+0.07}$ & $0.07_{-0.04}^{+0.02}$ & $-0.60_{-0.72}^{+0.49}$ & $0.00_{-0.03}^{+0.03}$ \\

~  & ~  & ~  & ~  & ~  & ~  & ~  & ~  & ~  \\

$\sigma_{int}$ & $0.32_{-0.04}^{+0.06}$ & $0.33_{-0.04}^{+0.05}$ & $0.28_{-0.05}^{+0.04}$ & $0.26_{-0.04}^{+0.04}$ & $0.23_{-0.03}^{+0.05}$ & $0.23_{-0.04}^{+0.04}$ & $0.08_{-0.04}^{+0.04}$ & $0.15_{-0.04}^{+0.06}$  \\
\hline
\end{tabular}
\end{center}
\label{tab: corrcoeffmb}
\end{table*}

The mean, median and $68\%$ confidence range read\footnote{Note that we exclude from this analysis the four galaxies from the Simon et al. (2005) sample since we do not have an estimate of their disc scalelength.}\,:

\begin{displaymath}
\langle \alpha_d \rangle = -1.03 \ , \ \hat{\alpha}_d = -1.02 \ , \ 68\% \ CL = (-1.52, -0.51) \ ,
\end{displaymath}

\begin{displaymath}
\langle \alpha_{0.1} \rangle = -2.28 \ , \ \hat{\alpha}_{0.1} = -2.33 \ , \ 68\% \ CL = (-2.63, -1.76) \ ,
\end{displaymath}
with both the distributions being almost symmetric around the mean values. Considering that $|\alpha(r)|$ approaches the null value in the very inner regions $(r << R_{-2})$ and that, for a typical $R_{vir}/R_d \sim 60$ value, $\alpha_{0.1}$ refers to the slope of the density profile in the terminal part of the rotation curve, it is easy to qualitatively understand why the Einasto model turns out to be so successful in fitting the data. Indeed, the dark haloes has a very shallow density profile in the disk dominated region $(r << R_d)$ thus mimicking cored models which are well known to fit the dwarfs galaxies rotation curve data. In the intermediate $r \sim R_d$ regions, the logarithmic slope takes similar values to the one of the \cite{B95} model which is known to work well in this regime. Finally, the Einasto model mimics the isothermal one (which has a constant logarithmic slope $\alpha = -2$) over the $r >> R_d$ range probed by the data. It is, therefore, this ability to interpolate among different models which makes the Einasto profile so successful in reproducing the rotation curve data of both dwarfs and normal spiral galaxies.

The quite large confidence ranges for $(n_{DM}, \alpha_d, \alpha_{0.1})$ may also be considered as an evidence against the existence of a universal halo profile. It is then worth investigating whether the environment takes some role in driving this non universality. Should this be the case, we must find a correlation between the environment and slope indicators. The correlation coefficient and the $(a, b, \sigma_{int})$ parameters of the fitted linear relations (using $\alpha_i/\hat{\alpha}_i$ with $i = d, 0.1$ as variables) are summarized in Table\,\ref{tab: corrcoeffslope}. Although ${\cal{C}}(x, y)$ can be sometimes large (as for the $\alpha_{0.1}$\,-\,$\Theta$ case) or the slope $a$ being not consistent with the null value within the $68\%$ CL (as for the $\alpha_d$\,-\,$D_{750}$ case) but with a large scatter, we can safely conclude that there is not any statistical evidence of a correlation of the logarithmic slopes with the environment. As an example, we plot $n_{DM}$ vs the environment indicators in Fig.\,\ref{fig: ndmplot} giving a visual confirmation of the former conclusion.

The same conclusion is strongly suggested by the results in Table\,\ref{tab: corrcoeffhalopar} for the concentration parameter and its correlation with the environment obtained by fitting $\log{(c_{vir}/\hat{c}_{vir})}$ vs $(\Theta, \log{\Sigma_{10}}, \log{D_{750}})$. It is particularly instructive to note that, although ${\cal{C}}(\Theta, c_{vir})$ is quite large hence arguing in favour of a strong correlation, the slope $a$, on the contrary, definitely shows that such a correlation does not exist at all. Such an example, therefore, highlights the importance of fully taking into account the errors on the quantities before drawing any conclusion on a relation between two quantities.

Somewhat surprisingly, we find a strong correlation among the virial mass and the local and global density estimators $(\Sigma_{10}, D_{750})$, but not with the tidal index $\Theta$. Contrary to what happens for $c_{vir}$, the ${\cal{C}}(x, y)$ values are always quite small, but the slopes $a$ of the correlations\footnote{Since $\mu_{vir}$ is already a logarithmic variable, one could use $\mu_{vir}/\hat{\mu}_{vir}$ as fitted quantity. We rather prefer to use $\mu_{vir} - \hat{\mu}_{vir} = \log{(M_{vir}/\hat{M}_{vir})}$ with $\hat{M}_{vir} = {\rm dex}(\hat{\mu}_{vir}) \simeq 3.98 \times 10^{10} \ {\rm M}_{\odot}$.} with $(\log{\Sigma_{10}}, \log{D_{750}})$ turn out to be non vanishing at the $68\%$ CL thus arguing in favour of a dependence of $\mu_{vir}$ on the environment (although with a non negligible scatter). It is worth noting that both $\Sigma_{10}$ and $D_{750}$ are not related to the dynamical state of the environment and can be contaminated by projection effects. As such, one can not exclude that the found correlation is only an artifact. Should we take them for actual relations, we should conclude that the higher is the local density $\Sigma_{10}$, the less massive is the dwarf halo. On the contrary, the higher is the global density $D_{750}$, the larger is the halo mass. Reconciling these two somewhat opposite behaviour is actually quite difficult so that we warn the reader to not overrate the significance of the detected $\mu_{vir}$\,-\,$\Sigma_{10}$ and $\mu_{vir}$\,-\,$D_{750}$ correlations.

\subsection{Dark to stellar ratios}

In order to shed some light on the interplay between the baryonic component and the dark halo of galaxies, it is interesting to investigate how they contribute to the total matter distribution. We have therefore estimated the DM mass fraction $f_{DM}(r) = M_{DM}(r)/[M_{bar}(r) + M_{DM}(r)]$ where we define an effective total baryon mass as $M_{bar}(r) = r v_{bar}^2{r}/G$ with $v_{bar}(r)$ the sum of the disk and gas (if any) circular velocities. Although such a definition strictly holds only for a spherical system and is therefore rigorously wrong for a thin disc, we prefer to use it since it can be easily related to the observed circular velocity data.

As for the logarithmic slope, we must choose a reference radius to estimate the DM mass fraction. We again consider two different possibilities, namely the disc scalelength $R_d$ and the halo characteristic radius $R_{-2}$. We then find\,:

\begin{displaymath}
\langle f_d \rangle = 0.74 \ , \ \hat{f}_d = 0.77 \ , \ 68\% \ CL = (0.61, 0.89) \ ,
\end{displaymath}

\begin{displaymath}
\langle f_{-2} \rangle = 0.75 \ , \ \hat{f}_{-2} = 0.85 \ , \ 68\% \ CL = (0.73, 0.91) \ .
\end{displaymath}
having defined $f_{d} = f_{DM}(R_d)$ and $f_{-2} = f_{DM}(R_{-2})$. Consistent with the common expectation, it turns out that dwarfs galaxies are DM dominated already at the disk scalelength radius $R_d$ (and hence still more at $R_{-2} > R_d$). Although such a result is model dependent, having been obtained fitting a specific halo model, the agreement of that model predictions with the measurements makes us confident that the above estimates of $(f_d, f_{-2})$ are fully realistic thus confirming the popular picture of dwarfs as ideal laboratories to study the dark halo properties.

Since all the dwarfs in our sample are DM dominated notwithstanding their different properties, we do not expect to find a correlation of $(f_d, f_{-2})$ with $(\Theta, \Sigma_{10}, D_{750})$. Table\,\ref{tab: corrcoefffd} shows that both the Spearman correlation coefficient ${\cal{C}}(x, y)$ and the slope $a$ of the $\log{(f_i/\hat{f}_i)}$ vs $(\Theta, \log{\Sigma_{10}}, \log{D_{750}})$ are strong evidence for the lack of correlation between the DM mass content and the environment the dwarf galaxy lives in. Actually, according to the $68\%$ CL range for $a$, one could argue in favour of a correlation between $f_d$ and the global density estimator $D_{750}$, but a look at the data in Fig.\,\ref{fig: fdmplot} shows that the points are quite scattered around the best fit line and with large errors so that better quality data are needed to confirm this preliminary and discrepant result.

An alternative way to look at the interplay between DM halo and baryons is to consider the ratio among a characteristic halo size and the disc one. We therefore estimate the two quantities ${\cal{R}}_{-2} = R_{-2}/R_d$ (which determines when the galaxy starts be fully DM dominated) and ${\cal{R}}_{vir} = R_{vir}/R_d$ (which is related to the total mass ratio $M_{vir}/M_d$). We find\,:

\begin{displaymath}
\langle {\cal{R}}_{-2} \rangle = 5.31 \ , \ \hat{{\cal{R}}}_{-2} = 3.99 \ , \ 68\% \ CL = (1.64, 16.75) \ ,
\end{displaymath}

\begin{displaymath}
\langle {\cal{R}}_{vir} \rangle = 65.0 \ , \ \hat{{\cal{R}}}_{vir} = 56.2 \ , \ 68\% \ CL = (18.8, 166.3) \ ,
\end{displaymath}
both distributions being characterized by long tails towards values larger than the mean ones. Fitting $\log{({\cal{R}}_i/\hat{{\cal{R}}}_i)}$ vs the environment estimators, we find the results summarized in Table\,\ref{tab: corrcoeffrd}. The slope $a$ of the fitted relations is always well consistent with $a = 0$ within the $68\%$ CL so that we can safely conclude that there is no impact of the environment on the size ratios. It is also interesting to note that the scatter in $\log{({\cal{R}}_{vir}/\hat{{\cal{R}}}_vir)}$ is the same from one case to another suggesting that this quantity is roughly constant along the sample. Since ${\cal{R}}_{vir} \propto (M_{vir}/M_d)^{1/3}$, such a result could be anticipated noting that, on such large scale, it is reasonable to expect $M_{vir}/M_d \propto \varepsilon \Omega_{CDM}/\Omega_b$ with $\Omega_{CDM}$ ($\Omega_b$) the CDM (baryon) cosmological density parameters and $\varepsilon$ a scaling factor quantifying the efficiency of the conversion of baryons into stars and HI/H$\alpha$ gas.

\subsection{Newtonian accelerations}

A quantity that has recently attracted a lot of interest is the Newtonian acceleration $g(r) = GM(r)/r^2$ where $M(r)$ is the mass of the DM halo or the baryon component. We reconsider here this issue for our dwarfs sample choosing the disc scalelength $R_d$ the halo scale radius $R_{-2}$ as reference positions. Setting $r = R_d$, we find\,:

\begin{displaymath}
\langle \gamma_{d}^{DM} \rangle = -8.55 \ , \ \hat{\gamma}_{d}^{DM} = -8.48 \ , \ 68\% \ CL = (-8.88, -8.22) \ ,
\end{displaymath}

\begin{displaymath}
\langle \gamma_{d}^{bar} \rangle = -9.09 \ , \ \hat{\gamma}_{d}^{bar} = -9.10 \ , \ 68\% \ CL = (-9.49, -8.70) \ ,
\end{displaymath}
where we have used the notation $\gamma_{i}^{j} = \log{g_{j}(r_i)}$ with $i = d$ ($i = -2$) for $r_i = R_d$ ($r_i = R_{-2}$) and $j = DM$ ($j = bar$) for the DM (baryon) Newtonian acceleration (measured in $cm/s^2$). For $r = R_{-2}$, we get\,:

\begin{displaymath}
\langle \gamma_{-2}^{DM} \rangle = -8.62 \ , \ \hat{\gamma}_{-2}^{DM} = -8.62 \ , \ 68\% \ CL = (-8.89, -8.29) \ ,
\end{displaymath}

\begin{displaymath}
\langle \gamma_{-2}^{bar} \rangle = -9.37 \ , \ \hat{\gamma}_{-2}^{bar} = -9.29 \ , \ 68\% \ CL = (-9.74, -9.09) \ .
\end{displaymath}
An issue hotly debated concerns whether it is possible to define a universal quantity evaluating the DM or baryon acceleration at a given reference radius. Should this be the case, the proposed quantity should take a constant value independent on the environment the galaxy lives in. We examine this issue by fitting $\log{[g_{j}(r_i)/\hat{g}_{j}(r_i)]}$ as a function of $(\Theta, \log{\Sigma_{10}}, \log{D_{750}})$ for the different $ij$ combinations we have considered\footnote{Note that we define $\hat{g}_{j}(r_i) = {\rm dex}(\hat{\gamma}_{i}^{j})$ with ${\rm dex}(x) = 10^x$.}. Tables\,\ref{tab: corrcoeffgd} and \ref{tab: corrcoeffgs} summarize the results of these fits. For almost all cases, the slope $a$ of the fitted relation is consistent with a null value within the $68\%$ CL indicating the absence of any statistical meaningful correlation. The only exception is represented by the $\gamma_{d}^{bar}$ which seems to correlate with both the tidal index $\Theta$ and the global density estimator $D_{750}$, but surprisingly not with the local density estimator $\Sigma_{10}$. The intrinsic scatter is, however, quite large and the constraints on $a$ too weak to definitely argue in favour of an impact of environment on the baryon acceleration at the disc scalelength radius.

The lack of correlations with the environment and the not too large $68\%$ confidence ranges may then be considered as evidence that the considered Newtonian accelerations are indeed roughly universal quantities. In order to further investigate this issue, we perform another series of fits taking as $x$ variable the absolute $B$ band magnitude ${\cal{M}}_B$ and the logarithm of the disc scalelength $\log{R_d}$ obtaining the results summarized in Table\,\ref{tab: corrcoeffmb}. It turns out that the baryon acceleration at $R_d$ and the DM one at $R_{-2}$ significantly correlates with the disc size so that they can not be considered as universal quantity. On the contrary, such a role could be played by the DM acceleration at $R_d$ and the Newtonian one at $R_{-2}$ since we have found no correlations with either the environment or the galaxy properties. It is worth noting, however, that the scatter around the median for both quantities is not negligible so that one should take this preliminary conclusion {\it cum grano salis}.

\section{Conclusions}

Being likely dark matter dominated, dwarfs and LSB galaxies are ideal targets to investigate the halo density profile in the inner regions. We have here used a large sample of such systems with well measured and radially extended rotation curves to constrain the dark matter halo parameters adopting the Einasto model for the density profile. Such a model has been proposed as a good description of dark matter haloes in high resolution N\,-\,body simulations and its viability has been recently confirmed by its success in fitting rotation curves of the THINGS galaxy sample \citep{CdBM11}. Our result strengthens these findings and allows to extend the applicability of the Einasto model to the dwarfs galaxies regime which has been not tested by either simulations or observations. In particular, we find that the Einasto slope parameter $n_{DM}$ can take values in a wide range which overlaps the one predicted by N\,-\,body simulations, but with a strong preference for shallower profiles. Whether this is due to the different mass range probed by dwarfs or a consequence of the impact of baryon collapse (not taken into account in the simulations) is still an open question which deserves further investigation.

The wide range spanned by $n_{DM}$ translates in a similarly large range for the logarithmic slope $\alpha(r)$ when evaluated at the disc scalelength radius $R_d$ or at a halo characteristic radius $0.1 R_{vir}$. As we have yet discussed, in order to fit the data, the parameters of the Einasto model are adjusted in such a way to mimic an almost cored profile in the very inner regions and an isothermal or NFW\,-\,like in the outer regions. How this transition takes place depends on the details of the galaxy so that we get a large range for both $\alpha_d$ and $\alpha_{0.1})$. We can therefore safely conclude that, contrary to what was argued by the past N\,-\,body simulations, there is no observational evidence for any universal halo profile.

Motivated by this result, one can then wonder what is driving the diversity of the halo profiles. In an attempt to get some hints on this question, we have here investigated the impact of the environment the galaxy lives in on the DM properties. To this end, we have resorted to three different indicators (namely, the tidal index $\Theta$, the local overdensity $\Sigma_{10}$ and the global overdensity $D_{750}$) to quantify whether a galaxy is isolated or reside in a rich environment. We have then used a Bayesian motivated procedure to fit the Einasto parameters and other DM related quantities as loglinear functions of $(\Theta, \Sigma_{10}, D_{750})$ noting that a null value for the slope of the fitted relation is a strong evidence for the absence of any correlation. According to this criterium, we can argue that the environment does not play any role in determining the final properties of the dark matter haloes of dwarfs galaxies. Actually, this is somewhat counterintuitive since, e.g., one can easily think of tidal interactions playing an important role in modifying the initial density profile of the dwarf halo. Should this be the case, we must find a difference between the halo parameters according to the properties of the environment. Since this is not the case, one should look for a different mechanism which is environment independent and able to wash out the impact of tidal interactions on the dark matter haloes.

Further work is actually needed in order to strengthen the preliminary conclusions we have presented here. First, the galaxy sample should be enlarged in order to better constrain the investigated correlations and estimate their intrinsic scatter. Although an updated version of the CNG catalog is in preparation (I. Karachentsev, private communication), it is actually hard to significantly increase the number of galaxies with both an available well measured rotation curve and a tidal index determination. As an alternative way, one could try to refine the local and global overdensity estimators in order to reduce the impact of chance projection and make them more closely related to the dynamical state of the galaxy. Should the combination of an updated sample and better environment indicators be available and still confirm the absence of any correlation, one should then open a new path towards understanding the physical processes driving the formation of dark matter haloes and the interactions between the dark and luminous galactic components.

\section*{Acknowledgments}

We warmly thank W.J.G. de Blok, S. Oh, J. Simon and R. Swaters for making their data available in electronic format. I.D. Karachentsev and V.E. Karachentseva are greatly acknowledged for discussion on the use of the tidal index and other environment estimators. We are also grateful to an anonymous referee for his/her comments that have significantly helped to improve the paper. This research has made use of the NASA/IPAC Extragalactic Database (NED) which is operated by the Jet Propulsion Laboratory, California Institute of Technology, under contract with the National Areonautics and Space Administration. VFC is funded by the Italian Space Agency (ASI).

\appendix

\section{Testing the method}

The analysis presented above relies on the implicit assumption that our fitting procedure correctly recovers the halo model parameters $(n_{DM}, c_{vir}, \log{M_{vir}})$ and hence the dark matter related quantities of interest we have discussed above. Although our method is a standard one, it is still not guaranteed a priori that the data probe the rotation curve with sufficient accuracy and sampling. Moreover, since the data only probe a limited radial range $(R_{min}, R_{max})$, it is possible that that $R_{min}/R_d$ ($R_{max}/R_d$) is too large (too small) to guarantee that we are correctly recovering the inner slope (the halo mass). It is therefore important to test whether our inferred constraints are biased or not.

To this end, we apply our fitting procedure to a set of simulated rotation curves generated according to the steps schematically sketched below.

\begin{enumerate}

\item{We choose a reference galaxy from the sample used in the text and set the disc $(M_d, R_d)$ and gas $(M_{HI}, R_{HI})$ parameters of the simulated galaxy as $(\epsilon_{md} M_d, \epsilon_{rd} R_d, \epsilon_{mg} M_{HI}, \epsilon_{rg} R_{HI})$ with $\epsilon_i$ randomly generated from uniform distributions in the $(0.1, 10.1)$ range for the masses and $(0.5, 2.5)$ for the scale radii.} \\

\item{Concerning the halo model parameters, we first set $n_{DM}$ by randomly extracting from a uniform distribution over the range $(1, 11)$. We then set the virial mass $M_{vir}$ solving the empirically motivated relation\,:

    \begin{displaymath}
    \left ( \frac{M_{HI}}{9 \times 10^9 \ {\rm M_{\odot}}} \right ) =
    \frac{\left [ M_{vir}/(9.5 \times 10^{11} \ {\rm M_{\odot}}) \right ]^{0.33}}{1 + \left [ M_{vir}/(9.5 \times 10^{11} \ {\rm M_{\odot}}) \right ]^{-0.77}} \ .
    \end{displaymath}
Finally, we set the concentration $c_{vir}$ by asking that the halo contribution to the total circular velocity at $R_d$ equals $\nu_{DM}$ with $\nu_{DM}$ randomly sampled from the range $(0.05, 0.95)$. It is worth stressing that this procedure allows us to simulate a wide range of different disc/halo combinations, from disc to halo dominated galaxies. Moreover, the halo parameters are fully realistic being in agreement with both the result of numerical simulations and actual observations.} \\

\item{Having set the input baryons and halo parameters, we sample the theoretical rotation curve over the range $(\varepsilon_{min} R_{d}, \varepsilon_{max} R_{d})$ with $0.1 \le \varepsilon_{min} \le 0.6$ and $2.0 \le \varepsilon_{max} \le 12.0$ so that we can mimic different radial range coverage. We divide this range in ${\cal{N}}_{sim}$ equally spaced bins and take one random point from each bin with ${\cal{N}}_{sim}$ the total number of points in the simulated curve chosen to be roughly equal to the one in the reference curve. } \\

\item{We finally assign to each $R$ in the list generated above a value of the total circular velocity extracted from a Gaussian distribution centred on the theoretical value and with a variance set to $0.5\%$. To this fake observed $v_c(R)$, we attach a measurement error which is similar to the one on the closest $R$ point in the reference rotation curve.} \\

\end{enumerate}
As reference galaxy, we use NGC\,3274, UGC\,4173 and UGC\,10310 and generate $\sim 50$ simulated rotation curves for each case thus getting a good statistics to check the efficiency of our fitting procedure. For each parameter $x$, we look at the distributions of $\Delta x/\sigma_x$ with $\Delta x = x_{inp} - x_{out}$ and $(x_{out}, \sigma_x)$ the median and (symmetrized) $68\%$ uncertainty as inferred from the fit. Note that, because of the scatter introduced in the simulated data, we do not expect to perfectly recover the input values (i.e., $\Delta x = 0$). Rather, we can consider the fitting procedure as reliable if the $\Delta x/\sigma_x$ distribution is centred on a value smaller than 1. This is indeed the case for all the parameters we have considered in the analysis presented in the paper. We can briefly summarize the results of this test as follows.

\begin{itemize}

\item{The halo model parameters $(n_{DM}, c_{vir}, \log{M_{vir}})$ and the stellar $M/L$ ratio are well recovered with the median values of $\Delta x/\sigma_x \sim (0.2, 0.3, 0.3, 0.1)$ and rms values smaller than 1 for all of them. Not surprisingly, we find that $\Delta x/\sigma_x$ is anticorrelated with $R_{max}/R_d$ for $x = (c_{vir}, \log{M_{vir}})$, i.e., the larger is radial range covered, the better are the concentration and virial mass recovered.} \\

\item{The logarithmic slope and the dark matter mass fraction at the disc scalelength are excellently recovered with typical $\Delta x/\sigma_x$ values close to null for most of the cases. Significantly larger values are obtained when the same quantities are estimated at the halo radius $R_{-2}$ as a consequence of the uncertainties in recovering both the concentration and the virial mass. We, however, stress that the median $\Delta x/\sigma_x$ values are still smaller than unity so that we can rely on the estimated values used in the paper.} \\

\item{The Newtonian accelerations are correctly recovered although with larger $\Delta_x/\sigma_x$ values. We have, however, checked that there is not any statistically meaningful correlation of $\Delta x/\sigma_x$ with any of the parameters entering the correlations investigated so that we are confident that the estimated quantities are not systematically biased.} \\

\end{itemize}
Motivated by these comforting results, we therefore safely conclude that our fitting procedure and the available data are sufficient to reliably estimate the halo parameters and the DM related quantities we have been interested in.

\end{document}